\documentclass[a4paper,12pt]{article}
\usepackage[a4paper,text={16.8cm,22.4cm}]{geometry}
\usepackage{epsfig,amsmath,amssymb,verbatim,colordvi,cite}
\usepackage[bookmarks=false]{hyperref}
\allowdisplaybreaks 
\newcommand{\beq}{\begin{equation}}
\newcommand{\eeq}{\end{equation}}
\newcommand{\bea}{\begin{eqnarray}}
\newcommand{\eea}{\end{eqnarray}}

\newcommand{\eq}[1]{(\ref{#1})}

\newcommand{\Sec}[1]{Section \ref{#1}}

\newcommand{\ie}{{\it i.e.}}
\newcommand{\GeV}{{\rm GeV}}
\newcommand{\TeV}{{\rm TeV}}

\newcommand{\Br}{{\rm BR}}

\let\oldmarginpar\marginpar
\renewcommand\marginpar[1]{\-\oldmarginpar[\raggedleft\scriptsize\sf
#1]{\raggedright\scriptsize\sf #1}} 

\begin{document}

\begin{titlepage}

\begin{flushright}
CERN-PH-TH/2012-270
\end{flushright}

\vspace{0.2cm}
\begin{center}
\Large\bf
MSSM: Cornered and Correlated
\end{center}

\vspace{0.2cm}
\begin{center}
Ulrich~Haisch$^{\hspace{0.25mm} 1}$ and Farvah~Mahmoudi$^{\hspace{0.5mm} 2,3}$\\

\vspace{0.4cm}
{\sl 
${}^1\hspace{0.25mm}$Rudolf Peierls Centre for Theoretical Physics, University of Oxford, \\
OX1 3PN Oxford, United Kingdom \\
${}^2\hspace{0.25mm}$CERN Theory Division, Physics Department, CH-1211 Geneva 23, Switzerland \\
${}^3\hspace{0.25mm}$Clermont Universit\'e, Universit\'e Blaise Pascal, CNRS/IN2P3, LPC, BP 10448, \\ 63000 Clermont-Ferrand, France
}
\end{center}

\vspace{0.2cm}
\begin{abstract}
\vspace{0.2cm}
\noindent 
Inspired by the latest results of ATLAS and CMS on the search for the standard model~(SM) Higgs scalar, we discuss in this article the correlations between Higgs-boson properties, low-energy observables, such as $B\to X_s \gamma$, $B_s\to \mu^+\mu^-$, and $(g-2)_\mu$, and the dark matter (DM) relic density. We focus on the corners of the MSSM parameter space where the $pp \to h \to \gamma \gamma$ signal is enhanced due to the presence of a light stau state. In this region $\tan\beta$, $M_A$, $A_t$, and $\mu$ take large values, and we find striking correlations between many of the considered observables. In particular, the $B\to X_s \gamma$ branching fraction is  enhanced, while the $B_s\to \mu^+\mu^-$ rate tends to be below the SM expectation.  In contrast, the Higgs-boson couplings show good overall agreement with the preliminary experimental determinations, the DM abundance is consistent with observation, and the discrepancy in $(g-2)_\mu$ is reduced. The predicted deviations and found correlations could be tested in the near future and hence may become very valuable as guidelines and consistency checks.
\end{abstract}
\vfil


\end{titlepage}

\section{Introduction} 
\label{sec:intro}

Recently both the ATLAS~\cite{ATLAS:2012gk} and the CMS~\cite{CMS:2012gu} collaborations have announced the existence of a new bosonic state with a mass of around  $125 \, \GeV$.  This discovery defines a turning point in the history of elementary-particle physics: the almost  50 year-long hunt for the Higgs boson has come to a dazzling end. One of the most central problems in high-energy physics is now whether the properties of the newly observed resonance agree with that of the standard model~(SM) Higgs scalar.  In fact, a major  experimental effort is directed towards shedding light on this question by measuring the various decay rates of the new particle as accurately as possible. While the achieved  precision is so far insufficient to draw any final conclusion, one cannot help but noticing that the LHC data match well the SM prediction with one possible exception:  the central value of the measured relative  strength of the diphoton signal  is too large by about~$70\%$~\cite{ATLAS:2012gk, CMS:2012gu}. If this enhancement survives further experimental scrutiny it may become the first convincing evidence of physics beyond the SM. 

An economic possibility to modify the effective coupling between the Higgs boson and two photons without altering its  main decay modes is provided by  extra vector-like leptonic states (see for example  \cite{Carena:2011aa,Carena:2012gp,Carena:2012xa, Ke:2012qc,Joglekar:2012vc,ArkaniHamed:2012kq,Almeida:2012he,Giudice:2012pf,Kearney:2012zi,Ajaib:2012eb,Bae:2012ir,Voloshin:2012tv,McKeen:2012he}). In the context of the minimal supersymmetric SM (MSSM) a light, maximally mixed stau  can be shown to increase the  loop-induced $h\gamma\gamma$ coupling~\cite{Carena:2011aa}, hence leading to the desired effect. 

In this article we dissect the ``light stau scenario'' extending previous  analyses~\cite{Carena:2011aa,Carena:2012gp,Ke:2012qc, Giudice:2012pf,Ajaib:2012eb,Arbey:2012dq} that are similar in spirit. In particular, we will show that enhancements of the $pp \to h \to \gamma \gamma$ signal associated to virtual stau exchange occur only in a narrow sliver in MSSM parameter space that features large (and positive) values of $\tan \beta$, the trilinear coupling $A_t$, and the $\mu$ parameter. In this corner of phase space we investigate in detail the indirect constraints arising from the most relevant $B$-physics observables, the muon anomalous magnetic moment ($(g-2)_\mu$), and dark matter (DM).  We observe striking and testable correlations between the individual observables. Our main findings are as follows: $i)$~An enhancement of the diphoton signal of the wanted order necessarily leads to an increase of the $B \to X_s \gamma$ branching ratio of around $30\%$. With improved theoretical and experimental determinations of the inclusive radiative $B$ decay this would represent an unambiguous signature of the light stau scenario. $ii)$~For positive gluino mass parameters $M_3$, the decay rate of $B_s \to \mu^+ \mu^-$ tends to be below its SM prediction with the exact value depending sensitively on $A_t$ and $\mu$. Future precision measurements of the purely leptonic $B_s$ decay by ATLAS, CMS, and LHCb hence represent sensitive probes of the MSSM parameters and its particle spectrum. $iii)$ The supersymmetric (SUSY) parameters that play an important role in in the $B$-physics observables also leave an imprint in the $h \to b \bar b$ decay rate which typically turns out to be larger than expected from a SM Higgs boson. Tentatively, this seems to be in accordance with the findings reported by CDF~\cite{Aaltonen:2012if} and D\O~\cite{Abazov:2012tf}. The  $h \to \tau^+ \tau^-$ channel, on the other hand, is in general suppressed, which is again in line with observation~\cite{CMS:2012gu}. $iv)$~Under the assumption of a light slepton/sneutrino spectrum the predicted MSSM corrections in $(g-2)_\mu$ allow for a good description of the experimental data. $v)$~In the presence of a light bino-like neutralino the specific MSSM scenario under study is compatible with thermal DM. In fact, the interplay between an enhancement in $h \to \gamma \gamma$ and the requirements imposed by the relic density essentially fixes the mass splitting between the lightest neutralino and stau. 

The intriguing correlations found in our article can be used as guidelines for the direct searches of SUSY particles, and will become extremely valuable as consistency and cross checks, in case the LHC high-$p_T$ experiments will start to see the first scalar partners.  

This work is divided into two parts. In Section~\ref{sec:results} we pursue an analytic approach which will give clear insights into the anatomy of the Higgs-boson properties, the low-energy observables, and the DM relic abundance in the region of MSSM parameter space that predicts a significantly enhanced diphoton signal.  Our analytic analyses will be complemented in Section~\ref{sec:numerics} by a detailed numerical study of the light stau scenario, employing state-of-the-art computer codes. A summary of our main results and conclusions are presented in Section~\ref{sec:conclusions}. 

\section{Analytic Results}
\label{sec:results}

In this section we present simple formulas that allow to understand the impact of the various MSSM parameters on the prediction of the Higgs-boson properties (its mass, production cross section, and decay rates) as well as on the expectations for the most important low-energy observables (\ie, $B \to X_s \gamma$, $B_s \to \mu^+ \mu^-$, and $(g-2)_\mu$). The requirements to achieve a proper thermal DM relic density are also discussed. Since in our numerical analysis of the MSSM parameter space, presented in \Sec{sec:numerics}, we will employ the most advanced calculations, including all  relevant contributions,  the analytic expressions presented hereafter  serve mostly an illustrative purpose.
  
\subsection{Anatomy of Higgs-Boson Mass}
\label{sec:higgsmass}

In the decoupling limit of the MSSM, \ie, $M_A^2 \gg M_Z^2$, the lightest CP-even Higgs boson  (for a review including original references, see  \cite{Djouadi:2005gj}) acquires the squared tree-level mass 
\beq \label{eq:mh2tree}
M_h^2 \approx M_Z^2 \, c^2_{2\beta} \left ( 1 - \frac{M_Z^2}{M_A^2} \, s_{2 \beta}^2 \right )\,.
\eeq
Here and in what follows we employ the shorthand notations $s_{2 \beta} = \sin \left ( 2 \beta \right)$, $c_{2 \beta} = \cos \left ( 2 \beta \right)$, {\it etc.}  Because $c_{2 \beta}^2 \propto 1$  in the large-$t_\beta$ limit and given the minus sign of the decoupling correction proportional  to $M_Z^2/M_A^2$, it follows  that $M_h^2 \leq M_Z^2$. Yet, the scalar field $h$ has SM-like couplings when the pseudo-scalar Higgs-boson mass $M_A$ is large, so that this state should have been discovered  at LEP, if it were not for the radiative corrections which push its mass upward from the tree-level upper bound of $M_Z$ to $M_h > 114.4 \, {\rm GeV}$ \cite{Barate:2003sz}. 

In fact, these higher-order corrections can be very large, since the scalar sector of the MSSM involves strong couplings, such as those to the top quark and its scalar partners the stops. In the limits $M_A, t_{\beta} \to  \infty$, that are the relevant ones for the upper bound on $M_h$, these corrections are  simple to evaluate. The dominant one-loop contribution to~\eq{eq:mh2tree}  arises from an incomplete cancellation of  top-quark and top-squark loops \cite{Okada:1990vk, Ellis:1990nz,Haber:1990aw}, and can be approximated by \cite{Casas:1994us, Dabelstein:1994hb,Carena:1995bx, Carena:1995wu, Haber:1996fp}
\bea \label{eq:Deltamhstop}
(\Delta M_h^2)_{\tilde t} \approx  \frac{3  G_F}{\sqrt{2} \pi^2}\, m_t^4 \left [ -L_{t \tilde t}+ \frac{X_t^2}{m_{\tilde t}^2} \left ( 1 - \frac{X_t^2}{12 m_{\tilde t}^2} \right )  \right ] \,, \hspace{2mm}
\eea
where  $L_{t \tilde t} = \ln \hspace{0.25mm} \big (m_t^2/m_{\tilde t}^2 \big )$ with  $m_{\tilde t}^2 = m_{\tilde t_1} m_{\tilde t_2}$, while $X_t = A_t - \mu/t_\beta$  denotes the stop-mixing parameter, which  depends on the trilinear stop-Higgs boson coupling $A_t$ and the higgsino mass parameter $\mu$.  We infer that for fixed stop spectrum scale $m_{\tilde t}$, the  Higgs-boson mass  correction from top/stop loops is maximised for  $|X_t|  = \sqrt{6} \hspace{0.5mm} m_{\tilde t} \approx 2.4 \hspace{0.5mm} m_{\tilde t}$, which is referred to as ``maximal mixing" scenario. The manifest symmetry of~\eq{eq:Deltamhstop} under the sign flip $X_t \to -X_t$ is broken by  finite  two-loop threshold corrections, which induce a term $4 \alpha_s/\pi  X_t/m_{\tilde t}$~\cite{Heinemeyer:1999be, Carena:2000dp} that leads to slightly larger values of the Higgs-boson mass for $X_t M_{3} \approx A_t M_{3}  > 0$. 

For large $t_\beta$ there are further  contributions to~\eq{eq:mh2tree} that can be relevant  \cite{Carena:1995bx, Carena:1995wu,Haber:1996fp}. These corrections arise from the sbottom and stau sector and take the form ($\tilde f = \tilde b, \tilde \tau$) 
\beq \label{eq:Dmh2sb}
(\Delta M_h^2)_{\tilde f} \approx -\frac{N_c^{\tilde f}}{\sqrt{2} G_F} \, \frac{y_f^4}{96 \pi^2} \, \frac{\mu^4}{m_{\tilde f}^4}  \,, 
\eeq
where  $N_c^{\tilde b} = 3$, $N_c^{\tilde \tau} =1$, $m_{\tilde f}^2 = m_{\tilde f_1} m_{\tilde f_2}$, and  we have ignored logarithmic  terms for simplicity. In the limit of interest, the bottom (tau) Yukawa coupling receives important one-loop correction whose dominant contribution depends on  ${\rm sgn} \left (\mu M_{3} \right )$ (${\rm sgn} \left (\mu M_2 \right )$) \cite{Hempfling:1993kv, Hall:1993gn, Carena:1994bv, Pierce:1996zz}.  The choice  $\mu M_{3} > 0$ ($\mu M_2 >0$) tends to reduce the tree-level Yukawa coupling $y_b = \sqrt{2} m_b/(v \hspace{0.125mm} c_\beta)$ ($y_\tau =   \sqrt{2} m_\tau/(v  \hspace{0.125mm} c_\beta)$), and as a result decreases the strictly negative sbottom (stau) effect~\eq{eq:Dmh2sb} on the Higgs-boson mass.

The relations~\eq{eq:mh2tree} to (\ref{eq:Dmh2sb}) allow to draw some general conclusions concerning the impact of the observation of a Higgs boson with $M_h \approx 125 \, \GeV$ on the MSSM parameter space. First, from the expression $M_h^2$, we deduce that  large values of  $t_\beta$ and $M_A$ are crucial to raise the  tree-level Higgs-boson mass as much as possible. In order to achieve a shift of $\Delta M_h^2 \approx (85.5 \, \GeV)^2$, the top/stop effects~\eq{eq:Deltamhstop} have to be very large, which typically requires $m_{\tilde t} \gtrsim 1 \, {\rm TeV}$ (heavy stop spectrum) and/or $|A_t| \gtrsim 2 \, {\rm TeV}$ (large stop mixing).  Since~\eq{eq:Dmh2sb} scales with the fourth power of $\mu$, the value of the higgsino mass parameter should not be too large. Finally, for ${\rm sgn} \left ( A_t M_{3} \right )$, ${\rm sgn} \left ( \mu M_{3} \right )$, and ${\rm sgn} \left (\mu M_2 \right)$  equal to $+1$,  subleading negative corrections to $M_h^2$ are minimised. We will see below that (some of) the mentioned MSSM parameters also play an important role in the production and the decay of the Higgs boson. This fact leads to interesting correlations. 

\subsection{Anatomy of Higgs-Boson Production and Decay}
\label{sec:higgsproductionanddecay}

An elegant way of obtaining the interactions of a light Higgs boson, is to construct an effective Lagrangian by integrating out heavy degrees of freedom. The resulting effective Higgs-boson couplings can be found most easily by utilising low-energy theorems, which relate amplitudes with different numbers of zero-momentum Higgs-boson fields~\cite{Spira:1995rr, Kniehl:1995tn} (see \cite{Ellis:1975ap,Shifman:1979eb} for the original idea). Here we will apply this general framework to illustrate the main features of Higgs-boson production and decay in the MSSM.  In order to present transparent formulas, we will again focus on the leading corrections in the decoupling limit~\cite{Djouadi:1996pb}. 

In the SM, Higgs-boson production via gluon-gluon fusion receives its  dominant contribution from triangle diagrams involving  top quarks. In the limit of a light Higgs boson, in which we are interested in, the corresponding form factor can be replaced by its  asymptotic value 
\beq \label{eq:topformfactor}
F_{1/2} (\tau_t) \approx \lim_{\tau_i \to \infty} F_{1/2} (\tau_i) = 1\,,
\eeq
where $\tau_i = 4 m_i^2/M_h^2$. In the infinite mass limit, one has furthermore \cite{Djouadi:2005gj}
\beq \label{eq:F0oF12}
\lim_{\tau_i \to \infty} \frac{F_{0} (\tau_i)}{F_{1/2} (\tau_i)} = \frac{1}{4}\,. 
\eeq
with $F_0 (\tau_i)$ encoding the effects of scalar loops. 

In the MSSM, the modification of the Higgs-boson production cross section in $gg \to h$, can be approximated by 
\beq \label{eq:Rh}
    R_h  = \frac{\sigma(gg\to h)_{\rm MSSM}}{\sigma(gg\to h)_{\rm SM}} = (1 + \kappa_g)^2 \approx \bigg ( 1 + \sum_{i = \tilde t, \tilde b} \kappa_i \bigg )^2  \,,
\eeq
where $\kappa_{\tilde t}$ and $\kappa_{\tilde b}$ represents the effects of top-squark and bottom-squark triangles, respectively. Notice that~\eq{eq:Rh} ignores the fact that the top-quark and bottom-quark Yukawa couplings in the MSSM differ from those in the SM. As will become clear later on, this mismatch is subleading in the $M_Z^2/M_A^2$ expansion, and hence can be neglected for our purposes.

The power of the Higgs low-energy theorems arises from the fact that in the decoupling limit the corrections $\kappa_{\tilde t, \tilde b}$ can be simply obtained by  differentiating the mass-squared matrices ${\cal M}_{\tilde t, \tilde b}^2$ with respect to the mass of the corresponding SM quark. For the top squarks, one has 
\beq \label{eq:stopmass}
{\cal M}_{\tilde t}^2 = \begin{pmatrix} \tilde m_{Q_3}^2 + m_t^2 + D_{Q_3} & m_t X_t \\ m_t    X_t & \tilde m_{u_3}^2 + m_t^2 + D_{u_3} \end{pmatrix} \,,
\eeq
where $\tilde m_{Q_3, u_3}^2$ are soft SUSY-breaking masses and $D_{Q_3,u_3} = {\cal O} (M_Z^2)$. A similar expression holds in the case of the bottom-squark sector.  Ignoring the contributions from the $D$-terms, which are numerically subleading, the master formula for the top-squark contribution to (\ref{eq:Rh}) reads
\beq \label{eq:kappagstop}
  \kappa_{\tilde t}  \approx \frac{1}{4} \, m_t^2 \, \frac{\partial}{\partial m_t^2} \, \ln \left [ \det \left ( {\cal M}_{\tilde t}^2 \right ) \right ]  \approx \frac{m_t^2}{4} \left ( \frac{1}{m_{\tilde t_1}^2} + \frac{1}{m_{\tilde t_2}^2} - \frac{X_t^2}{m_{\tilde t_1}^2 m_{\tilde t_2}^2} \right )   \,,
\eeq
where the multiplicative factor $1/4$ in the first line stems from the normalisation (\ref{eq:F0oF12}) of the scalar form factor,  and the final result agrees with the expression given in~\cite{Dermisek:2007fi, Cacciapaglia:2009ky}. From~(\ref{eq:kappagstop}) one infers that the amount of mixing in the stop sector, parametrised by $X_t$, determines whether the ratio~\eq{eq:Rh} is smaller or larger than 1. For no mixing, corresponding to $X_t = 0$, one has $R_h \geq 1$, so that Higgs-boson production in $gg \to h$ is enhanced with respect to the SM. On the other hand, if $X_t$ is parametrically larger than the mass eigenvalues $m_{\tilde t_{1,2}}$  (with  $m_{\tilde t_1} \leq m_{\tilde t_2}$) of (\ref{eq:stopmass}), then $R_h \leq 1$, meaning that the Higgs boson  is less likely to be produced. The fact that in the MSSM, in order to make the Higgs boson sufficiently heavy, one needs large/maximal mixing, {\it i.e.}, $|X_t| \approx \sqrt{6 \hspace{0.5mm} m_{\tilde t_1}m_{\tilde t_2}}$, then tells us that for a random MSSM parameter point that gives $M_h \approx 125 \, {\rm GeV}$ one should find a suppression of $\sigma (gg \to h)$. In fact, this is precisely what happens. As a final remark, we add that the sign of the new-physics corrections to Higgs-boson production in  gluon-gluon fusion is in many models closely related to the (non-)cancellation of the quadratic divergence in the Higgs-boson mass \cite{Low:2009di}. From~\eq{eq:Deltamhstop} and~\eq{eq:kappagstop}, we see that in the MSSM there is a strong anti-correlation between $M_h$ and $R_h$, driven by $X_t$. 

The calculation of the sbottom contribution to $R_h$ proceeds along the lines of~\eq{eq:kappagstop}. Since $m_b^2 \ll m_t^2$,   notable effects can only arise if the mixing in the sbottom sector is very large. In this limit, one has approximately 
\beq \label{eq:kappagsbottom}
  \kappa_{\tilde b}  \approx -\frac{m_b^2 X_b^2}{4 m_{\tilde b_1}^2 m_{\tilde b_2}^2}   \,,
\eeq
where $X_b = A_b - \mu \, t_\beta$. Obviously, this correction is strictly destructive and can only be important if either the trilinear term $A_b$ and/or the combination $\mu \, t_\beta$ is sufficiently larger than the sbottom masses $m_{\tilde b_{1,2}}$. 

In order to describe the decays of the Higgs boson, we define the corrections factors 
\beq \label{eq:Gammahtof}
\frac{\Gamma (h \to VV)_{\rm MSSM}}{\Gamma (h \to VV)_{\rm SM}} = ( 1 + \kappa_V )^2 \,, \qquad \frac{\Gamma (h \to f \bar f)_{\rm MSSM}}{\Gamma (h \to f \bar f)_{\rm SM}}  = ( 1 + \kappa_f )^2 \,,
\eeq
for $V = W, Z, \gamma$ and $f = b, \tau, t$. 

With respect to~\eq{eq:kappagstop}, the derivation of the leading non-decoupling corrections to $\kappa_{V,f}$ is complicated~\cite{Spira:1995rr, Kniehl:1995tn, Cacciapaglia:2009ky} by the fact that in the MSSM one has not a single, but two neutral scalar fields which develop a vacuum expectation value (VEV), 
\beq \label{eq:HuHd}
H_u = \frac{1}{\sqrt{2}} \left ( v_u + c_\alpha  h   +  s_\alpha  H  \right ) \,,  \qquad H_d = \frac{1}{\sqrt{2}} \left ( v_d - s_\alpha  h + c_\alpha H \right ) \,.
\eeq
Here ($-\pi/2 \leq \alpha \leq 0$)
\beq \label{eq:alpha}
\alpha = \frac{1}{2}  \arctan \left ( t_{2 \beta} \, \frac{M_A^2 + M_Z^2 }{M_A^2 - M_Z^2 } \right ) \,,
\eeq
and $h$ ($H$) denotes the lighter (heavier) CP-even Higgs-boson mass eigenstate. Furthermore, $v_u/v_d = t_\beta$ and $\sqrt{v_u^2 + v_d^2}= v \approx 246 \, {\rm GeV}$, so that  $v = v_u/s_\beta = v_d/c_\beta$. 

In the case of $\kappa_W$, which encodes the modification of the Higgs-boson coupling to a pair of $W$ bosons, the presence of the mixing angle $\alpha$ in~\eq{eq:HuHd} leads to  
\beq \label{eq:kappaW}
\kappa_W = -1+ \frac{v}{M_W} \left ( c_\alpha \, \frac{\partial}{\partial v_u} - s_\alpha \, \frac{\partial}{\partial v_d}  \right) M_W  = - 1 - s_{\alpha - \beta}  \approx -\frac{M_Z^4}{8 M_A^4} \, s^2_{4 \beta} \,,
\eeq
where we have used that $M_W = g/2 \, \sqrt{v_u^2 + v_d^2}$ and in the last step expanded~\eq{eq:alpha} in powers of $M_Z^2/M_A^2$, retaining only the first non-zero term in the Taylor series. An analogue formula applies to $\kappa_Z$, which implies that the Higgs-boson couplings to massive gauge-boson pairs are affected in a universal way, $\kappa_V = \kappa_W = \kappa_Z$. This universal correction~\eq{eq:kappaW} is strictly destructive, but of order $M_Z^4/M_A^4$, and thus numerically insignificant as long as one sticks to the decoupling limit. Since $s^2 _{4 \beta} \propto 1/t_\beta^2$, it is further reduced in the large-$t_\beta$ limit. 

In the case of the Higgs-boson couplings to tau leptons only the second term of the differential operator in~\eq{eq:kappaW} contributes because $m_\tau = v_d/\sqrt{2} \, y_\tau$. One obtains 
\beq \label{eq:kappatau}
\kappa_\tau  = - 1 -  \frac{s_\alpha}{c_\beta} \approx  -\frac{2 M_Z^2}{M_A^2} \, s^2_\beta \,  c_{2\beta} \,.
\eeq
The correction $\kappa_t$ is obtained from~\eq{eq:kappatau} by the replacements $-s_\alpha \to c_\alpha$, $c_\beta \to s_\beta$, and $s_\beta^2 \to -c^2_\beta$. Since $m_t = v_u/\sqrt{2} \, y_t$ only the first term in the bracket of~\eq{eq:kappaW} results in a correction to $\kappa_t$. We see that the shifts in the tree-level couplings of the Higgs-boson to fermion pairs fall off quadratically in the limit $M_A^2 \gg M_Z^2$, and that $\kappa_{\tau} \geq 0$ whereas $\kappa_{t} \leq 0$.  In the large-$t_\beta$ limit $\kappa_{\tau} \propto 1$ and $\kappa_t \propto 1/t_\beta^2$. 

The coupling of the Higgs boson to bottom quarks receives important $t_\beta$-enhanced loop corrections involving charginos and gluinos \cite{Carena:1994bv,Carena:1999py}.  In the limit $t_\beta \gg 1$, we find including the leading terms 
\beq \label{eq:kappab}
\kappa_b  \approx  \frac{1}{1+\epsilon_b \hspace{0.25mm} t_\beta} \frac{M_h^2 + (\Delta M_h^2)_{\tilde t} + M_Z^2}{M_A^2}\,,
\eeq
where  the expression for the tree-level Higgs-boson mass and the dominant one-loop correction are given in (\ref{eq:mh2tree}) and~(\ref{eq:Deltamhstop}), respectively. Furthermore, 
\beq \label{eq:deltab}
\epsilon_b = \frac{\mu A_t }{16\pi^2} \frac{y_t^2}{m_{\tilde t}^2} \, f (x_{\tilde t \mu}) + \frac{2 \alpha_s}{3 \pi} \frac{\mu M_3}{m_{\tilde b}^2}  \, f(x_{\tilde b 3})\,
\eeq
with $x_{\tilde t \mu} = m_{\tilde t}^2/\mu^2$, $x_{\tilde b 3} = m_{\tilde b}^2/M_3^2$, and 
\beq \label{eq:ifun}
f (x) =-\frac{x}{1-x} - \frac{x}{(1-x)^2} \, \ln x \,.
\eeq
Notice that $f(x) $  is positive definite with $f(0) = 0$, $f(1) = 1/2$, and $f(\infty) = 1$.  This feature together with the appearance of the combinations $\mu A_t$ and $\mu M_3$ in (\ref{eq:deltab}) makes the correction $\kappa_b$ introduced in (\ref{eq:kappab}) quite sensitive to the choice of MSSM parameters. While  the modifications $\kappa_{b,t}$ have only a minor impact on Higgs-boson production, as already anticipated  in~\eq{eq:Rh}, we will  see below, that the decoupling corrections to $\kappa_{b}$  can  play an important role for the decays of a light Higgs boson. 

In order to give an explicit result for $\kappa_\gamma$, we recall that within the SM, the process $h \to \gamma \gamma$ is dominated  by the virtual loop-exchange of $W$ bosons. These contributions interfere destructively with the top-quark amplitude.  One has
\beq \label{eq:FW}
F_W = -\lim_{\tau_i \to \infty} \frac{F_{1} (\tau_W)}{F_{1/2} (\tau_i)} \approx 6.24 \,,
\eeq
where $F_1 (\tau_i)$  is the form factor associated with vector-boson loops~\cite{Djouadi:2005gj}  and the numerical result corresponds to our reference value $M_h = 125 \, \GeV$ for the Higgs-boson mass. With the help of $F_W$, we write the modification of the Higgs-boson coupling to two photons as 
\begin{eqnarray}\label{eq:kappagamma}
   \kappa_\gamma \approx  \frac{1}{F_W- \frac{4}{3}} \, \Bigg[  -\frac{4}{3} \, \kappa_{\tilde t}  -\frac{1}{3} \, \kappa_{\tilde b} -  \kappa_{\tilde \tau}  + \kappa_{H^\pm} +  \kappa_{\chi^\pm} \Bigg] , \hspace{2.5mm}
\end{eqnarray}
where the stop and sbottom   contributions, \ie, $\kappa_{\tilde t}$ and $\kappa_{\tilde b}$,  have already been given in~\eq{eq:kappagstop} and~\eq{eq:kappagsbottom}. Notice that compared to~\eq{eq:Rh} they  appear here with opposite signs, signalling that constructive interference in $gg \to h$ goes along with destructive interference in $h \to \gamma \gamma$ and {\it vice versa}. This correlation between the loop-induced  effective $hgg$ and $h\gamma \gamma$ couplings is a general feature in new-physics models with coloured fermionic/scalar partners. 

The diphoton channel also receives contributions from stau, charged Higgs-boson, and chargino loops. The stau corrections takes the form 
\beq \label{eq:kappagstau}
  \kappa_{\tilde \tau}  \approx -\frac{m_\tau^2 X_\tau^2}{4 m_{\tilde \tau_1}^2 m_{\tilde \tau_2}^2}   \,,
\eeq
with  $X_\tau = A_\tau - \mu \, t_\beta$. Like $\kappa_{\tilde b}$ the correction~\eq{eq:kappagstau} is only important if the stau-mixing parameter satisfies $X_\tau \gg m_{\tilde \tau_{1,2}}$ and  the lighter stau  mass eigenstate is not too heavy. The former  requirement demands both  $t_\beta$ and $\mu$ to be large, and in this region of parameter space, stau loops necessarily lead to an enhancement of $\Gamma (h \to \gamma \gamma)$. 

The charged Higgs-boson effects are, on the other hand, strictly destructive in the MSSM. In the decoupling limit, we find
\beq \label{eq:kappaahiggs}
  \kappa_{H^\pm} \approx -\frac{1}{4} \left ( M_W^2 - \frac{1}{2} \hspace{0.25mm} M_Z^2  \hspace{0.25mm}  c_{2\beta}^2 \right)  \frac{\partial}{\partial M_W^2}   \ln \left ( M_{H^\pm}^2 \right ) = -\frac{2 M_W^2 -  M_Z^2  \hspace{0.25mm}  c_{2\beta}^2}{8 M_{H^\pm}^2} \,,
\eeq
where $M_{H^\pm}^2 = M_A^2 + M_W^2$. Because the spin-zero amplitude $F_0 (\tau_i)$ is suppressed by a factor of $1/4$ relative to $F_{1/2}  (\tau_i)$ (see~\eq{eq:F0oF12}), and $M_{H^\pm}^2 \approx M_A^2$ in the decoupling limit, the correction~\eq{eq:kappaahiggs} has only a very minor effect on the diphoton decay. Note that $c_{2\beta}^2 \to 1$ for $t_\beta \to \infty$. 

The last ingredient in~\eq{eq:kappagamma} is provided by triangle graphs with internal chargino exchange. In terms of the chargino mass matrix 
\beq \label{eq:charginomass}
{\cal M}_{\chi^\pm} = \begin{pmatrix} M_2 &  \sqrt{2} M_W s_\beta \\ \sqrt{2} M_W c_\beta &\mu \end{pmatrix} \,,
\eeq
the corresponding coefficient can be written as 
\beq \label{eq:kappaachargino}
  \kappa_{\chi^\pm}  \approx - M_W^2 \, \frac{\partial}{\partial M_W^2} \, \ln \left [ \det \left (  {\cal M}_{\chi^\pm}^T {\cal M}_{\chi^\pm} \right ) \right ]  \approx {\rm sgn} \left [ \det \left ( {\cal M}_{\chi^\pm} \right )  \right ] \, \frac{2 M_W^2}{m_{\chi_1^\pm} \, m_{\chi_2^\pm}} \, s_{2 \beta}  \,. 
\eeq
Since for sufficiently large values of $M_2$ and $\mu$, one has $\det \left ( {\cal M}_{\chi^\pm} \right ) \approx \mu M_2$, the overall sign of $\kappa_{\chi^\pm}$ is determined by the one of the product of the higgsino and wino mass parameters.  It follows that for $\mu M_2   > 0$ ($ \mu M_2  < 0$), charginos enhance (suppress) the $h \to \gamma \gamma$ rate. Since $s_{2 \beta} \propto 1/t_\beta$ the effects are largest for low $t_\beta$, and numerically very important if the chargino spectrum is light.

When converting the above results into branching ratios, one must bear in mind that the total decay rate $\Gamma(h)$ of a light SM-like Higgs boson  is dominated by its decay into bottom quarks. For  $M_h=125 \,{\rm GeV}$, one has $\Br(h\to b\bar b)\approx 60\%$, $\Br(h\to WW)\approx 21\%$, $\Br(h\to gg)\approx 7\%$, $\Br(h\to \tau^+ \tau^-)\approx 6\%$, and $\Br(h\to ZZ)\approx 3\%$. It then follows from~\eq{eq:Gammahtof} that 
\beq \label{eq:RGamma}
   R_\Gamma = \frac{\Gamma(h)_{\rm MSSM}}{\Gamma(h)_{\rm SM}}  
   \approx 0.60 \left ( 1 + \kappa_b \right )^2  + 0.07\left ( 1+\kappa_g \right) ^2 + 0.33 \,. 
\eeq
Note that  only the shift $\kappa_{b}$ has been included here, while the tree-level corrections $\kappa_\tau$ and $\kappa_{W,Z}$ have been neglected. This is a very good approximation, since $\Br(h\to b\bar b) \gg \Br(h\to \tau^+ \tau^-)$ and~\eq{eq:kappaW} is relative to~\eq{eq:kappab} suppressed by an additional power of $M_Z^2/M_A^2$.

At this point we are ready to work out the products of the production cross section times branching ratios for the various Higgs-boson decay channels. These are the key observables that will be affected by the different MSSM contributions. Defining
\beq \label{eq:Rf}
\begin{split}
   R_X & = \frac{\big[\sigma(pp\to h)\,\Br(h\to X)\big]_{\rm MSSM}}
              {\big[\sigma(pp\to h)\,\Br(h\to X)\big]_{\rm SM}} \\[1mm] & \approx \frac{1}{R_\Gamma} \prod_{i=g,X} (1+\kappa_i)^2  \approx 1 + 1.86 \, \kappa_g  - 1.20 \, \kappa_b + 2 \, 
              \kappa_X \,,    
\end{split}                     
\eeq
we  find the following semi-analytic results for the most interesting final states $X$ containing either massive vector bosons 
\beq \label{eq:RV}
R_V  \approx 1 + 0.47 \left ( \frac{m_t^2}{m_{\tilde t_1}^2} + \frac{m_t^2}{m_{\tilde t_2}^2} - \frac{m_t^2 X_t^2}{m_{\tilde t_1}^2 m_{\tilde t_2}^2} - \frac{m_b^2 X_b^2}{m_{\tilde b_1}^2 m_{\tilde b_2}^2} \right )  - 1.20 \, \frac{1}{1+\epsilon_b \hspace{0.25mm} t_\beta} \frac{M_h^2 + (\Delta M_h^2)_{\tilde t} + M_Z^2}{M_A^2} \,, 
\eeq
where $V=W,Z$ or diphotons
\beq \label{eq:Rgamma}
\begin{split}
R_\gamma & \approx 1 + 0.33  \left ( \frac{m_t^2}{m_{\tilde t_1}^2} + \frac{m_t^2}{m_{\tilde t_2}^2} - \frac{m_t^2 X_t^2}{m_{\tilde t_1}^2 m_{\tilde t_2}^2} \right) - 0.43 \, \frac{m_b^2 X_b^2}{m_{\tilde b_1}^2 m_{\tilde b_2}^2}  + 0.10 \, \frac{m_\tau^2 X_\tau^2}{m_{\tilde \tau_1}^2 m_{\tilde \tau_2}^2} \\[1mm]
&  \phantom{xx} + 1.63 \,  {\rm sgn} \left (\mu M_2 \right  ) \, \frac{M_W^2}{m_{\chi_1^\pm} \, m_{\chi_2^\pm}} \, \frac{1}{t_\beta}  - 1.20 \, \frac{1}{1+\epsilon_b \hspace{0.25mm} t_\beta} \frac{M_h^2 + (\Delta M_h^2)_{\tilde t} + M_Z^2}{M_A^2} \,.
\end{split}
\eeq
The result for the relative signal strength $R_b$ of the $b \bar b$ channel is obtained from (\ref{eq:RV}) by simply replacing $-1.20$ by $0.80$. In order to obtain the above expressions we have included all non-decoupling corrections, \ie,~\eq{eq:kappagstop},~\eq{eq:kappagsbottom},~\eq{eq:kappagstau}, and~\eq{eq:kappaachargino}, but apart from~\eq{eq:kappab} neglected all contributions that vanish in the limit $M_A^2 \gg M_Z^2$. We furthermore took the limit $t_\beta \to \infty$, keeping only the leading corrections, and for simplicity replaced ${\rm sgn} \left [ \det \left ( {\cal M}_{\chi^\pm} \right ) \right ]$ by ${\rm sgn} \left (\mu M_2 \right)$. We also remark that measurements of the double ratio $R_\gamma/R_{W,Z} \approx 1 + 2 \, \kappa_\gamma$ would allow for a clean extraction of the $h \to \gamma \gamma$ amplitude, since it is  independent of  $\kappa_{\tau,b}$ (see \cite{Djouadi:2012rh} for a recent detailed study).  

The formulas~\eq{eq:RV} and~\eq{eq:Rgamma} are the main result of this section. They  exhibit  interesting correlations with the expressions presented in \Sec{sec:higgsmass}. Focusing first on  the correction $(\Delta M_h^2)_{\tilde t}$ introduced in~\eq{eq:Deltamhstop}, we see that  in the limit of maximal stop mixing $|X_t| \approx \sqrt{6 \hspace{0.5mm} m_{\tilde t_1}m_{\tilde t_2}}$ (needed to lift the Higgs-boson mass to $125 \, \GeV$) the ratios $R_{W,Z}$ and $R_\gamma$ are necessarily reduced. In terms of $m_{\tilde t}^2 = m_{\tilde t_{1}} m_{\tilde t_{2}}$, the shift in $R_{W,Z}$ ($R_\gamma$) is given approximately by $-1.9 \, m_t^2/m_{\tilde t}^2$ ($-1.3 \, m_t^2/m_{\tilde t}^2$), which amounts to a correction of around $-5\%$ ($-4\%$) for $m_{\tilde t} = 1 \, {\rm TeV}$. Large mixing in the bottom-squark sector will further suppress the latter ratios. The decoupling corrections affecting the Higgs-boson couplings to the bottom and tau go in the same direction. Numerically, one finds a universal shift of $-2\%$ for $M_A = 1 \, {\rm TeV}$. Positive corrections to $R_\gamma$ can arise from chargino loops if ${\rm sgn} \left (\mu M_2 \right ) = +1$, which, as  mentioned below~\eq{eq:Dmh2sb}, helps also to diminish the negative correction $(\Delta M_h^2)_{\tilde \tau}$ to the Higgs-boson mass. Since~\eq{eq:kappaachargino} is $t_\beta$-suppressed,  there is however a generic tension between  large chargino effects in $R_\gamma$ and saturating  the upper limit on the tree-level Higgs-boson mass following from~\eq{eq:mh2tree}. A way to enhance $R_\gamma$ without running into immediate problems at the tree level, is provided by  a light stau with large mixing $X_\tau \approx \mu \, t_\beta$, which requires that both $t_\beta$ and $\mu$ are large. To give an example, employing $m_{\tilde \tau}^2 =  m_{\tilde \tau_1} m_{\tilde \tau_2} = (200 \, {\rm GeV})^2$, $t_\beta = 50$, and $\mu = 1 \, {\rm TeV}$ in~\eq{eq:Rgamma}, one finds that $R_\gamma$ is changed by $+50\%$. Since the correction $(\Delta M_h^2)_{\tilde b,\tilde \tau}$ in~\eq{eq:Dmh2sb} is proportional to $-\mu^4/ m_{\tilde b, \tilde \tau}^2$, one expects however an anti-correlation  between the size of the stau contribution to $R_\gamma$ and the loop-corrected Higgs-boson mass. As we will discuss in the next section, some of the MSSM corrections to  low-energy observables are also changed significantly in the limit $|\mu| \to \infty$, which leads to further correlations, strengthening the constraints on the parameter space. 

\subsection{Anatomy of Low-Energy Observables}
\label{sec:lowenergy}

The discussion in the previous two sections should have made clear that the part of the MSSM parameter space with $M_A, t_\beta \to \infty$ represents a phenomenologically interesting region for Higgs-boson physics. In the following, we will extract the dominant MSSM corrections to $B \to X_s \gamma$, $B_s \to \mu^+ \mu^-$, and $(g-2)_\mu$ in this limiting case, highlighting  the existing correlations with $M_h$, $R_b$, $R_{W,Z}$, and $R_\gamma$. 

We start our discussion by considering  the inclusive radiative $B \to X_s \gamma$ decay. Taking into account only the most important corrections due to the Wilson coefficient $C_7$ of the electromagnetic dipole operator  leads to the following ratio~\cite{Misiak:2006zs, Misiak:2006ab, Freitas:2008vh}
\beq \label{eq:RXs}
R_{X_s}  =  \frac{{\Br} (B \to X_s \gamma)_{\rm MSSM}}{{\Br} (B \to X_s \gamma)_{\rm SM}}  \approx 1- 2.61 \,  \Delta C_7 + 1.66 \, ( \Delta C_7 )^2 \,,
\eeq 
where $ \Delta C_7$ represents the additive correction appearing in the high-scale Wilson coefficient $C_7 = C_7^{\rm SM} +  \Delta C_7 \approx -0.19 + \Delta C_7$ \cite{Misiak:2004ew}.

Within the MSSM, the Wilson coefficient $\Delta C_7$ receives important contributions from loops involving tops and charged Higgs bosons, $\Delta C_7^{H^\pm}$, as well as stops and charginos,  $\Delta C_7^{\chi^\pm}$. In the decoupling limit with $t_\beta \gg 1$, the former corrections  can be approximated by 
\beq \label{eq:C7chargedHiggs}
\Delta C_7^{H^\pm} \approx \frac{m_t^2}{3 M_{H^\pm}^2} \left (L_{t H^\pm}  + \frac{3}{4} \right ) \,,
\eeq
where $L_{t H^\pm} = \ln \left (m_t^2/M_{H^\pm}^2 \right )$ with $M_{H^\pm}^2 \approx M_A^2$. Subleading $t_\beta$-enhanced corrections dominated by  gluinos \cite{Degrassi:2000qf, Carena:2000uj} have not been included here. Such effects enhance (suppress) $\Delta C_7^{H^\pm}$ for negative (positive) values of $\mu M_{3}$, but are irrelevant for our considerations. For $M_A = 1 \, {\rm TeV}$ one finds $\Delta C_7^{H^\pm} \approx -0.03$, where the minus sign reflects the well-known fact that in $B \to X_s \gamma$ the charged  Higgs-boson corrections interfere constructively with the SM amplitude \cite{Grinstein:1987pu, Hou:1987kf}.
 
 Since the charged Higgs-boson effects~\eq{eq:C7chargedHiggs} are not $t_\beta$ enhanced they render only an  insignificant correction in the  MSSM parameter region of interest. The dominant one-loop contributions to~\eq{eq:RXs} are therefore provided by diagrams with up-type squarks and charginos. These grow linearly with $t_\beta$ and take the form 
 \beq \label{eq:C7chargino}
\Delta C_7^{\chi^\pm} \approx - \mu A_t \, t_\beta \, \frac{m_t^2}{m_{\tilde t}^4} \, g (x_{\tilde t \mu}) \,.
\eeq
Here 
\beq \label{eq:f}
g (x) = -\frac{7 x^2-13 x^3}{12 \left (1-x \right )^3}- \frac{2 x^2 - 2 x^3 -3 x^4}{6 \left (1-x \right )^4}\, \ln x \,.
\eeq
Notice that we have included above only the correction due to top squarks and higgsino-like charginos, while the wino-like contribution has been suppressed. In the limit of large higgsino mass parameters, that we are mainly interested in,  this is a good approximation, because  the latter corrections  scale as $M_2/\mu \, M_W^2/m_{\tilde t}^2$ for $|\mu| \gg M_2$. The result (\ref{eq:C7chargino}) agrees with the expression given in \cite{Altmannshofer:2010zt}.  Since the function $g(x)$ is strictly positive  for $x \in \, ]0, \infty [$ with $g (1) = 5/72 \approx 0.07$, the sign in  \eq{eq:C7chargino} implies that for $\mu A_t > 0$ ($\mu A_t < 0$) the branching ratio is larger (smaller) than the  SM expectation \cite{Carena:1994bv, Barbieri:1993av}. The dominant $t_\beta$-enhanced gluino corrections,  not shown in \eq{eq:C7chargino}, again enhance (suppress)  the chargino contribution for  ${\rm sgn} \left (\mu M_{3} \right ) = -1 \; (+1)$, but leave the qualitative dependence of $\Delta C_7^{\chi^\pm} $ on $\mu A_t$ unchanged \cite{Degrassi:2000qf, Carena:2000uj}. For the choices $t_\beta = 50$, $m_{\tilde t} =  1.5 \, \TeV$, $|\mu|  = 1 \, \TeV$, and $|A_t| = 3 \, {\rm TeV}$, one finds $\Delta C_7^{\chi^\pm} \approx -{\rm sgn} \left ( \mu A_t \right ) \, 0.12$, which depending on the sign of $\mu A_t$ corresponds to an enhancement/suppression of the ratio~\eq{eq:RXs} by ${\cal O} (30\%)$. Shifts of this size in  ${\Br} (B \to X_s \gamma)$ are detectable given  the present theoretical calculations and experimental extractions. 

After $B \to X_s \gamma$, we now analyse the structure of the MSSM contributions to another ``standard candle" of quark flavour physics, namely $B_s \to \mu^+ \mu^-$.  We begin by defining the ratio
\beq \label{eq:Rmumu}
R_{\mu^+\mu^-}  =  \frac{{\Br} (B_s \to \mu^+ \mu^-)_{\rm MSSM}}{{\Br} (B_s \to \mu^+ \mu^-)_{\rm SM}} \approx  1 - 13.2  \; C_P + 43.6 \left (C_S^2 + C_P^2 \right )  \,,
\eeq
where $C_S$ and $C_P$ denote the dimensionless Wilson coefficients of the semileptonic scalar and pseudo-scalar operators.  In the large-$t_\beta$ regime these coefficients have the most important impact on $R_{\mu^+ \mu^-}$. The term linear in $C_P$ arises from the interference with the SM contribution  $C_{10}^{\rm SM} \approx -4.2$~\cite{Bobeth:1999mk, Bobeth:2003at} to the Wilson coefficient of the semileptonic axial-vector operator.  Importantly, for $C_P >0$ it interferes destructively with the  term proportional to $\left (C_S^2 + C_P^2 \right )$. This implies, on the one hand, that for positive values of $C_P$ the stringent bounds on $B_s \to \mu^+ \mu^-$ are more easily evaded and, on the other hand, that a pseudo-scalar contribution of the correct size will lead to a suppression of the purely leptonic decay mode below its SM value. 

Within the MSSM the contributions to  $C_{S}$ and $C_P$  with the strongest $t_\beta$ dependence arise from neutral Higgs double penguins  \cite{Babu:1999hn}.  In the decoupling limit, one has  $C_S \approx -C_P$ with 
\beq \label{eq:CP}
C_P \approx \mu A_t \, \frac{t_\beta^3}{(1+\epsilon_b \hspace{0.25mm} t_\beta)^2} \; \frac{m_t^2}{m_{\tilde t}^2} \, \frac{m_b  m_\mu}{4s_W^2 M_W^2 M_A^2} \, f (x_{\tilde t \mu})  \,.
\eeq
Henceforth we use the shorthand notation $s_W = \sin \theta_W$ {\it etc.} to indicate trigonometric functions of the weak mixing angle. Notice that our definition of the semileptonic scalar and pseudo-scalar Wilson coefficients differs from that of \cite{Bobeth:2001jm} by a factor of $m_b$. The parameter~$\epsilon_b$  introduced in (\ref{eq:deltab}), parametrises loop-induced non-holomorphic terms that receive their dominate contributions from higgsino and gluino exchange. The Wilson coefficients $C_S$ and $C_P$ also receive various other contributions  in the MSSM~\cite{Bobeth:2001jm,Chankowski:2000ng}, but these are of no concern as long as one is interested  in the general structure of the effects only. Recall that $f(x) >0$, so that the sign of $C_S$ ($C_P$) is opposite to (follows) that of the combination $\mu A_t$. Notice finally that   \eq{eq:deltab} introduced a dependence on ${\rm sgn} \left (\mu M_{3} \right )$ and that \eq{eq:CP} is suppressed (enhanced) for $\mu M_3 > 0$ ($\mu M_3 < 0$).

From what has been said below \eq{eq:Rmumu}, it should be clear, that the latter feature has important implications. In particular, it follows that finding ${\Br} (B_s \to \mu^+ \mu^-)$ close to its SM value leads to a two-sided bound on the product $\mu A_t$, if double Higgs-penguin contributions provide the dominant new-physics effect in the purely leptonic decay. For example, consider $R_{\mu^+ \mu^-} < 1.3$. Combining~\eq{eq:Rmumu} and~\eq{eq:CP},  translates into 
\beq \label{eq:Bsmmconstraint}
-\frac{0.16}{{\rm TeV}^2} \lesssim \frac{1}{(1+\epsilon_b \hspace{0.25mm} t_\beta)^2} \frac{\mu A_t}{m_{\tilde t}^2 M_A^2} \left ( \frac{t_\beta}{50} \right)^3 \lesssim \frac{1.37}{{\rm TeV}^2}\,.
\eeq
For simplicity we have set $x_{\tilde t \mu} =1$ here, which corresponds to the case of degenerate masses. Clearly, this double inequality  provides a non-trivial constraint on the MSSM parameter space with $t_\beta \gg 1$. Realize that the apparent asymmetry in \eq{eq:Bsmmconstraint} arises from the fact that  $C_P > 0$ is preferred over $C_P < 0$, because a small positive pseudo-scalar contribution leads to a cancellation between the linear and quadratic terms  in~\eq{eq:Rmumu}, and that values $1/(1+\epsilon_b \hspace{0.25mm} t_\beta)^2 \hspace{0.25mm} \mu A_t/(m_{\tilde t}^2 M_A^2) \, ( t_\beta/50 )^3 \approx 0.6/\TeV^2$ correspond to a reduction of ${\Br} (B_s \to \mu^+ \mu^-)$ relative to the SM  by about $50\%$. 

We now turn our attention to the anomalous magnetic moment $a_\mu = (g-2)_\mu/2$ of the muon.  In the MSSM, this observable receives one-loop contributions from Feynman diagrams with neutralino and smuon as well as chargino and sneutrino exchange (see \cite{Martin:2001st, Stockinger:2006zn} for topical reviews). The latter diagrams dominate in almost  the entire parameter space~\cite{Moroi:1995yh} and can be approximated by 
\bea \label{eq:amuchargino}
\Delta a_\mu^{\chi^\pm} \approx -\frac{G_F M_W^2}{\sqrt 2 \pi^2}\,  \mu M_2 \,  t_\beta  \, \frac{m_\mu^2}{m_{\tilde \nu}^4} \, \frac{h(x_{\chi_1 \tilde \nu})-h(x_{\chi_2 \tilde \nu})}{x_{\chi_1 \tilde \nu}-x_{\chi_2 \tilde \nu}} \,, \hspace{2.5mm}
\eea
with $x_{\chi_i \tilde \nu} = m_{\chi^\pm_i}^2/m_{\tilde \nu}^2$ and 
\beq \label{eq:h}
h (x) = \frac{x-3}{(1-x)^2}- \frac{2 }{(1-x)^3} \ln x\,.
\eeq
Since $x_{\chi_1 \tilde \nu} \leq x_{\chi_2 \tilde \nu}$ by convention and $h(x)$ is a monotonically decreasing function of $x$ with $h(1) = 2/3$ and $h(\infty) = 0$, it follows that the sign of \eq{eq:amuchargino} is determined by that of $\mu M_2$. Employing $t_\beta = 50$,  $m_{\tilde \nu} = |\mu| = 1\, \TeV$, $|M_2|=300 \,{\rm GeV}$, $m_{\chi^\pm_1} \approx |M_2|$, and $m_{\chi^\pm_2} \approx |\mu|$, one finds a shift $\Delta a_\mu^{\chi^\pm} \approx {\rm sgn} \left ( \mu M_2 \right ) 2.2 \cdot 10^{-9}$, which for $\mu M_2 > 0$ removes almost entirely the well-known tension between the experimental result and the SM prediction for $a_\mu$. 

The formulas \eq{eq:C7chargino}, \eq{eq:CP}, and  \eq{eq:amuchargino} provide a good starting point to discuss the correlations between the low-energy and Higgs-boson observables, which we expect to see  below in our numerical analysis. Choosing, for definiteness, ${\rm sgn} \left ( M_i \right ) = +1$ with $i = 1, 2, 3$, we infer from Sections~\ref{sec:higgsmass} and \ref{sec:higgsproductionanddecay} that the parameter region with large $t_\beta$ and large and positive  $A_t$ and $\mu$ is well suited, on the one hand, to explain the relative heaviness of the Higgs boson, and, on the other hand, to allow for a significant enhancement of the $h \to \gamma \gamma$ signal.  From \eq{eq:C7chargino} it follows that the above parameter choices inevitably  lead to an enhancement of the branching ratio of $B \to X_s \gamma$. While the $B_s \to \mu^+ \mu^-$ decay rate can in principle be both enhanced as well as reduced for $\mu A_t >0$, we will see in the next section, that for $\mu M_3 > 0$ one typically observes a suppression of the $B_s \to \mu^+ \mu^-$ branching ratio whenever the diphoton signal is  enhanced. Notice that the presence of the $t_\beta$-enhanced corrections $\epsilon_b$ in (\ref{eq:CP}) plays a crucial role in this context. In fact, from (\ref{eq:Bsmmconstraint}) it is evident that parameter choices that give $\epsilon_b> 0$ are favoured over those that lead to  $\epsilon_b< 0$. Finally, since ${\rm sgn} \left ( \mu \right ) = +1$ in the considered parameter region, one also expects a visible improvement in the description of the $a_\mu$ data,  if the sneutrinos are sufficiently light. 
  
\subsection{Anatomy of DM Relic Abundance}
\label{sec:darkmatter}

Another appealing feature of the special MSSM parameters under consideration, is the possibility to generate the correct DM relic density 
\beq \label{eq:relic}
\Omega_{\rm DM}  h^2 \approx \frac{1.07 \cdot 10^9}{\rm GeV} \frac{x_f}{M_{\rm Pl} \hspace{0.25mm}  \sqrt{g_\ast} \hspace{0.5mm} \hat \sigma_{\rm eff}} \,, 
\eeq 
through stau coannihilation  \cite{Griest:1990kh, Edsjo:1997bg, Ellis:1999mm, Gomez:1999dk} with the lightest SUSY particle (LSP), represented by the lightest neutralino mass eigenstate. In the above equation, $M_{\rm Pl} \approx 1.22 \cdot 10^{19} \, {\rm GeV}$ is the Planck mass, $g_\ast \approx 81$ is the  number of relativistic degrees of freedom at the freeze-out temperature $T_f$, and $x_f = m_{\chi_1^0}/T_f \approx 22$ for the masses of the lightest neutralino we are interested in. 

In the case at hand, the effective cross section 
\beq \label{eq:seff}
\hat \sigma_{\rm eff} \approx \alpha_{\chi \chi} \hspace{0.25mm} a_{\chi \chi} + \alpha_{\chi \tilde \tau} \hspace{0.25mm} a_{\chi  \tilde \tau} + {\cal O}\left (1/x_f \right ) \,,
\eeq
entering \eq{eq:relic} receives two important contributions. The first one arises from neutralino annihilation into tau pairs, induced by  stau exchange in the $t$ and $u$ channel. For a mostly bino-like neutralino, $m_{\chi_1^0} \approx |M_1|$, and large mixing in the stau sector, we find the expression
\beq \label{eq:achichi}
a_{\chi\chi} \approx \frac{e^4}{32 \pi c_W^4} \frac{1}{m_{\chi_1^0}^2} \, j (r_{\tilde \tau \chi}) \,, 
\eeq
where $r_{\tilde \tau \chi} = m_{\tilde \tau_1}/m_{\chi^0_1}$ and 
\beq \label{eq:j}
j(r) =  \frac{1}{( 1+ r^2)^2} \,.
\eeq
  
In order to achieve a pronounced $h \to \gamma \gamma$ signal, the  stau has to be  light and strongly coupled to the Higgs. It turns out that in this case the dominant coannihilation channel is $\chi_1^0 \tilde \tau_1 \to h \tau$, which proceeds through the $t$-channel exchange of the lighter stau state~\cite{Carena:2012gp}. Its contribution to~\eq{eq:seff} is approximately given by  
\beq \label{eq:achistau}
a_{\chi\tilde \tau} \approx \frac{5e^4}{1024 \pi s_W^2 c_W^4} \frac{m_\tau^2 \hspace{0.25mm} X_\tau^2}{M_Z^2  \hspace{0.25mm} m_{\chi_1^0}^4} \, k (r_{\tilde \tau \chi}, r_{h\chi}) \,, 
\eeq
with $r_{h\chi} =  m_{h}/m_{\chi^0_1}$ and 
\beq \label{eq:krs}
k(r_1,r_2) = \frac{(r_1-r_2+1)^2 (r_1+r_2+1)^2}{r_1 (r_1+1) \left(r_2^2-r_1 (r_1+1)^2\right)^2} \,.
\eeq
The results \eq{eq:achichi} and \eq{eq:achistau} can be derived from the general expressions given in \cite{Gomez:1999dk}. Notice that the coannihilation contribution $a_{\chi\tilde \tau}$ is proportional to $m_\tau^2 X_\tau^2$. Like in  \eq{eq:kappagstau} this factor arises from the $\tilde \tau_1 \tilde \tau_1 h$ coupling. 

The last missing ingredient needed to allow for a qualitative understanding of $\Omega_{\rm DM}  h^2$, are the Boltzmann weight factors $\alpha_{\chi \chi}$ and $\alpha_{\chi \tilde \tau}$ appearing in \eq{eq:seff}. In the parameter space that allows to reproduce the observed relic density, we obtain
\beq \label{eq:alphas}
\alpha_{\chi\chi}  \approx 1 \,, \qquad  \alpha_{\chi\tilde \tau}  \approx \frac{1}{2} \, e^{20.7 \, (1- r_{\tilde \tau \chi} )} \,.
\eeq
The expressions \eq{eq:alphas} show that the size of coannihilation is exponentially sensitive to the mass splitting $m_{\tilde \tau_1} - m_{\chi_1^0}$, meaning that the process $\chi_1^0 \tilde \tau_1 \to h \tau$ only becomes important in regions of parameter space that permit values of $r_{\tilde \tau \chi}$ close to 1. 

Based on \eq{eq:relic} to \eq{eq:alphas}, we are now able to discuss the restrictions that the observed relic density $\Omega_{\rm DM} h^2 \approx 0.11$ imposes on the MSSM parameter space. For neutralino masses $m_{\chi_1^0} \lesssim 100 \, {\rm GeV}$, it turns out that $\chi_1^0 \chi_1^0 \to \tau^+ \tau^-$ represents the dominant DM annihilation channel. In this case, one has $\Omega_{\rm DM} h^2 \approx (\Omega_{\rm DM} h^2)_{\chi \chi} $ with 
\beq \label{eq:Omegah2bino}
(\Omega_{\rm DM} h^2)_{\chi \chi} \approx 1.4 \cdot 10^{-2} \hspace{0.5mm} \left ( \frac{m_{\chi_1^0}}{0.1 \, {\rm TeV}} \right )^2 (1+ r_{\tilde \tau \chi}^2)^2 \,.
\eeq
It follows that in order to achieve a large enough DM  abundance one needs a sizeable mass splitting $m_{\tilde \tau_1} - m_{\chi_1^0}$. For example, for $m_{\chi^0_1} \approx 30 \, {\rm GeV}$ masses of $m_{\tilde \tau_1} \approx 90 \, {\rm GeV}$ result in viable $\Omega_{\rm DM} h^2$ values. Increasing the neutralino mass requires slightly  larger splittings $m_{\tilde \tau_1} - m_{\chi_1^0}$ that reach a maximum at $m_{\chi^0_1} \approx 60 \, {\rm GeV}$. At this point the mass difference starts decreasing, leading to $m_{\tilde \tau_1} - m_{\chi^0_1} \approx 35 \, {\rm GeV}$ for $m_{\chi^0_1} \approx 100 \, {\rm GeV}$. 

For somewhat heavier neutralinos $m_{\chi_1^0} \gtrsim 100 \, {\rm GeV}$ the proper DM density requires mass differences $m_{\tilde \tau_1} - m_{\chi^0_1}$ of  order $20 \, {\rm GeV}$, and as a result \eq{eq:achistau} has to be included. In consequence, $\Omega_{\rm DM} h^2 \approx (\Omega_{\rm DM} h^2)_{\chi \chi} + (\Omega_{\rm DM} h^2)_{\tilde \tau \chi}$ with the latter contribution approximated by  
\beq \label{eq:Omegah2stauchi}
(\Omega_{\rm DM} h^2)_{\tilde \tau \chi} \approx - 2.5 \left ( \frac{X_\tau}{50 \, \rm TeV} \right )^2 \, e^{20.7 \left ( 1- r_{\tilde \tau \chi} \right )} \,.
\eeq
Notice that the exponential suppression of $(\Omega_{\rm DM} h^2)_{\tilde \tau \chi}$ by $m_{\tilde \tau_1} - m_{\chi^0_1}$  is in the specific MSSM scenarios we consider balanced by the large mixing parameter $X_\tau \approx \mu \, t_\beta$. \\

\section{Numerical Analysis}
\label{sec:numerics}

\begin{figure}[!t]
\begin{center}
\includegraphics[width=0.32 \textwidth]{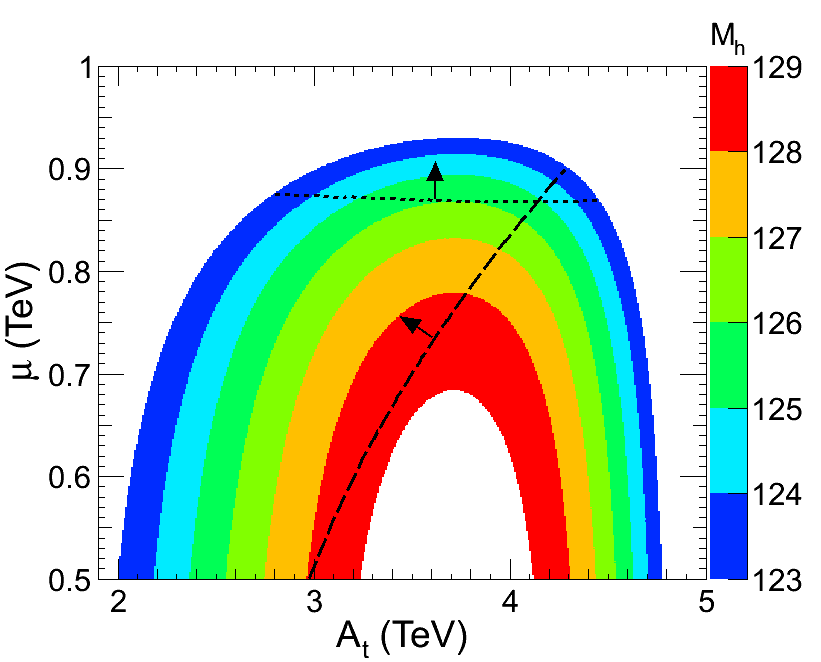}
\includegraphics[width=0.32 \textwidth]{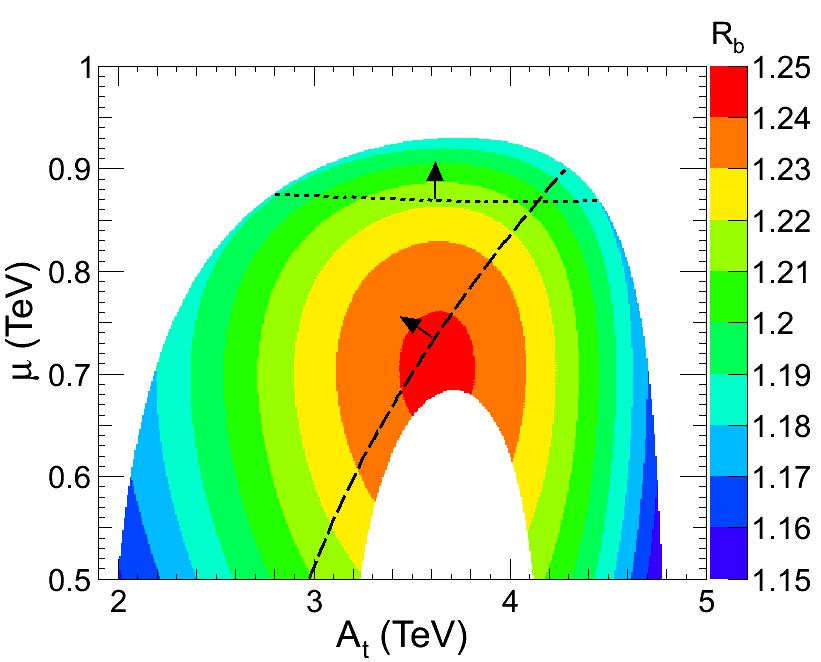}
\includegraphics[width=0.32 \textwidth]{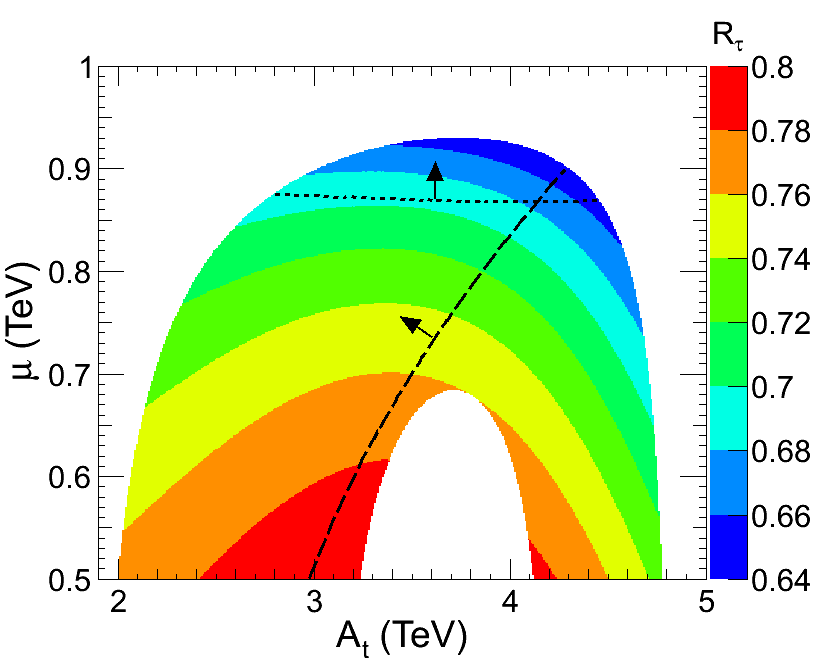}\\
\includegraphics[width=0.32 \textwidth]{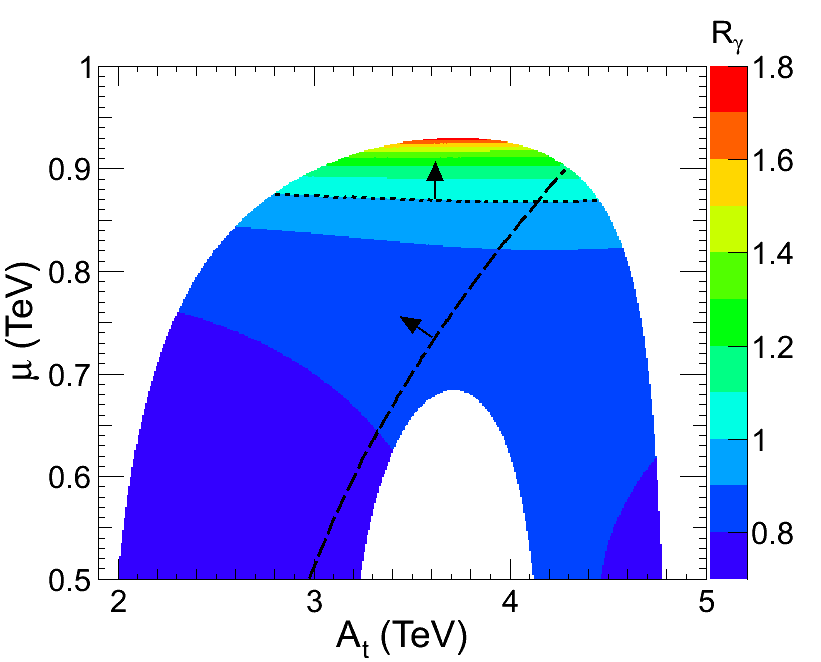}
\includegraphics[width=0.32 \textwidth]{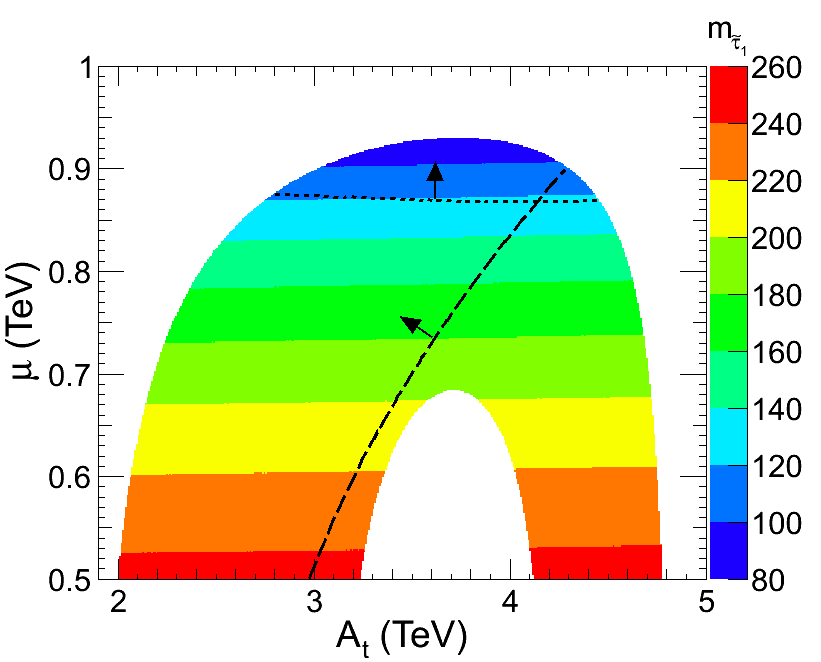}
\includegraphics[width=0.32 \textwidth]{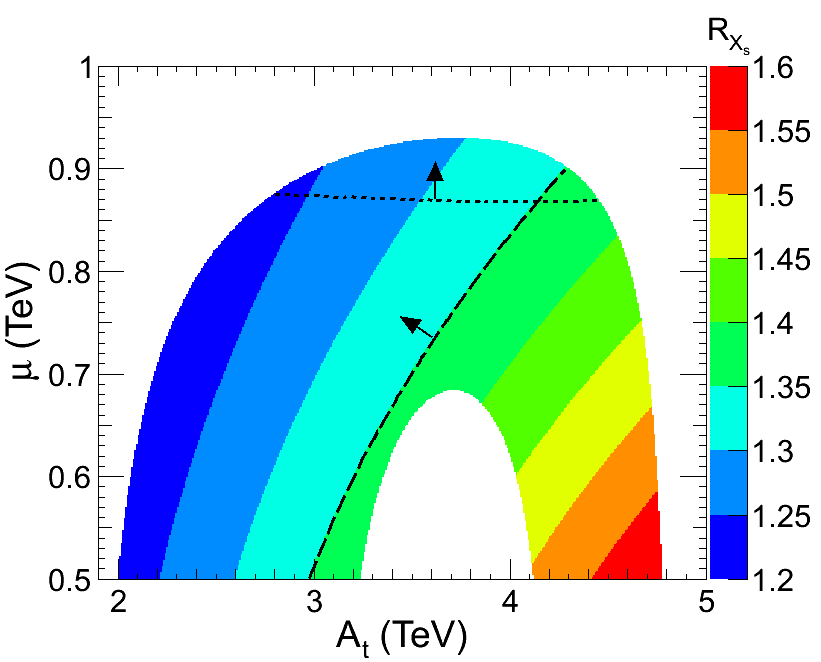}\\
\includegraphics[width=0.32 \textwidth]{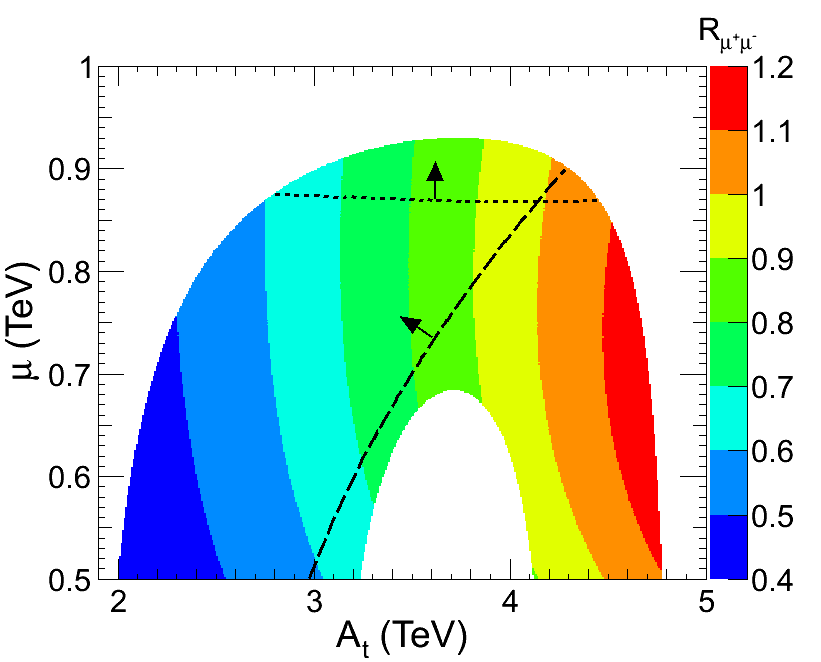}
\includegraphics[width=0.32 \textwidth]{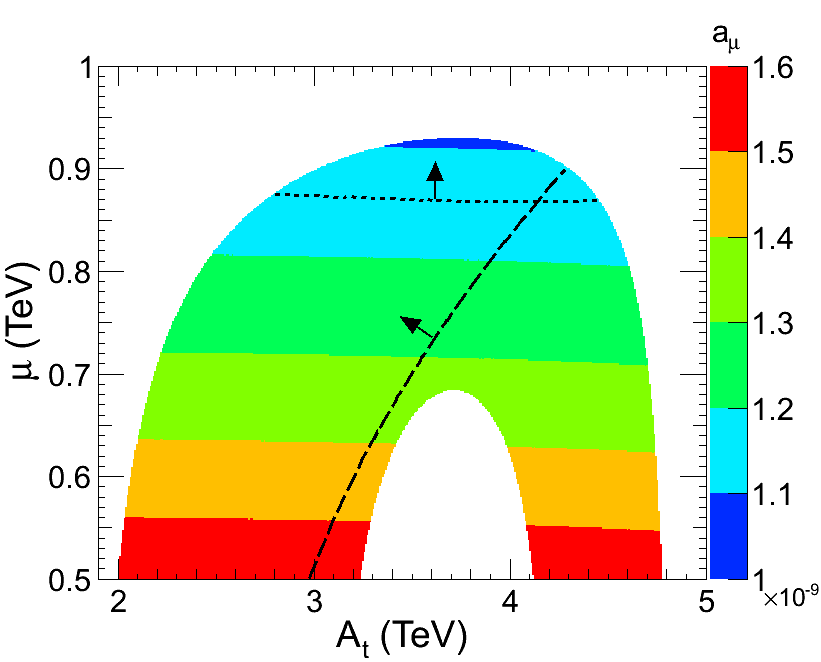}
\includegraphics[width=0.32 \textwidth]{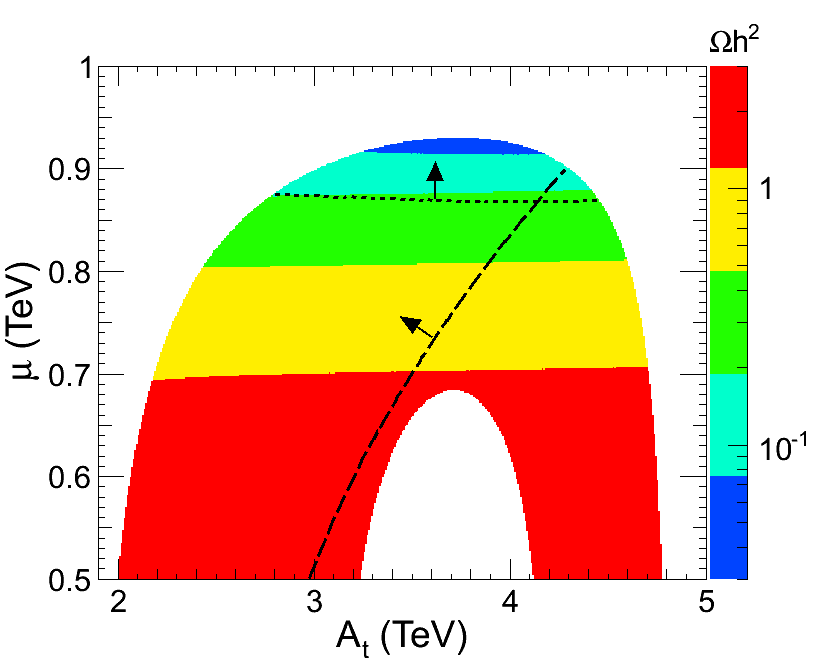}
\end{center}
\vspace{-6mm}
\begin{center}
\parbox{15.5cm}{\caption{{Results for $M_h$ (upper left), $R_{b}$ (upper center), $R_{\tau}$ (upper right), $R_{\gamma}$ (center left), $m_{{\tilde \tau}_1}$ (center), $R_{X_s}$ (center right), $R_{\mu^+ \mu^-}$ (lower left), $\Delta a_\mu$~(lower center), and $\Omega_{\rm DM} h^2$~(lower right) in the $t_\beta = 60$ scenario. The Higgs-boson and lighter stau masses are given in units of $\rm GeV$. The dotted black lines indicate the parameter regions with $R_\gamma > 1$, while the dashed black lines correspond to the $95\% \,{\rm CL}$ regions favoured by $B \to X_s \gamma$. See text for details.}\label{fig:Atmugrid}}}
\end{center}
\end{figure} 

We now turn to the numerical analysis of the Higgs-boson and  the low-energy observables as well as the DM relic density. All results presented below have been obtained using {\sc SoftSusy~3.3.3}~\cite{Allanach:2001kg} for the spectrum calculation, {\sc Higlu~3.11}~\cite{Spira:1995mt} and {\sc Hdecay~4.45}~\cite{Djouadi:1997yw} for the computation of Higgs-boson production and decay, and {\sc  SuperIso Relic~3.3}~\cite{Mahmoudi:2007vz,Arbey:2009gu} for the calculation of the relevant low-energy observables and the DM abundance.  We have compared  our results against {\sc SuSpect~2.41}~\cite{Djouadi:2002ze} and {\sc FeynHiggs~2.8.6}~\cite{Heinemeyer:1998yj}, and found good overall agreement between the different programs for most of the observables. The biggest numerical differences arise for the Higgs-boson mass and its branching ratios. In the former (latter) case we find relative changes of typically  below $5\%$ ($10\%$). The observed differences can be traced back, on the one hand, to the use of different renormalisation prescriptions ($\overline{\rm DR}$ {\it vs}. on-shell scheme) and, on the other hand, to the different treatment of higher-order perturbative corrections. The quoted relative errors  give  an indication of the theoretical uncertainty plaguing our calculations of the Higgs-boson observables, and we will comment on its impact on our numerical analysis below.  For a detailed comparisons between the publicly available programs dealing with the mass of the Higgs boson in the MSSM, we refer to \cite{Allanach:2004rh}. 

To begin with, we focus on scenarios with $t_\beta = 60$. The other choices of the MSSM parameters are $M_A = 1\, {\rm TeV}$, $M_1 = 50 \, {\rm GeV}$, $M_2 = 300 \, {\rm GeV}$, $M_3 = 1.2 \, {\rm TeV}$, $\tilde m_{Q_3} = \tilde m_{u_3} = 1.5 \, {\rm TeV}$, $\tilde m_{L_3} = \tilde m_{l_3} = 350 \, {\rm GeV}$, while we take common soft SUSY-breaking masses of $1.5 \, {\rm TeV}$ and $2 \,{\rm TeV}$ ($1 \, {\rm TeV}$) for the remaining ``left-handed" and ``right-handed" squark (sleptons). We furthermore employ  $A_b = 2.5 \, {\rm TeV}$ and $A_\tau = 500 \, {\rm GeV}$, while the first and second generation trilinear couplings take the same values as those of the third generation, as they have essentially no impact on the observables in question. Hereafter we will refer to this specific choice of parameters as the ``$t_\beta = 60$ scenario". In Figure~\ref{fig:Atmugrid} we show the results of our numerical scans in the $A_t$--$\mu$ plane. We restrict ourselves to the quadrant  with $A_t >0$ and $\mu > 0$, which shows the most interesting effects and correlations and is the only one that allows for a good description of all data. 

From the prediction for the Higgs-boson mass, we see that for the above choice of SUSY  parameters, the trilinear term $A_t$ has to lie in the range of $[2, 5] \, {\rm TeV}$ in order to push $M_h$ up to $[123, 129] \,{\rm GeV}$. The latter range is allowed by the ATLAS and CMS data~\cite{ATLAS:2012gk, CMS:2012gu}, if one accounts for the parametric errors from the SM input  (with the largest uncertainty arising from the  top-quark mass, $m_t^{\rm pole} = (173.3 \pm 2.8) \, {\rm GeV}$~\cite{Alekhin:2012py}) as well as the theoretical uncertainties in the MSSM calculation of the mass of the Higgs boson~\cite{Allanach:2004rh}. Of course, the need for large trilinear stop-Higgs boson couplings is an immediate consequence of~\eq{eq:mh2tree} and~\eq{eq:Deltamhstop}. The anti-correlation between the Higgs-boson mass and the $\mu$ parameter, as implied by~\eq{eq:Dmh2sb}, is also clearly visible in the panel. We emphasise that the shown predictions correspond to $\overline{\rm DR}$ input parameters and that the results obtained in the  on-shell scheme (as used for example in {\sc FeynHiggs}) would differ to some extent. In particular, the values for $A_t$ needed  to accommodate a Higgs-boson mass consistent with the LHC observations can be smaller by a factor of up to  $2$. This pronounced dependence on the renormalisation prescription should be kept in mind when interpreting our numerical results. 
 
We now turn our attention to Higgs-boson production and decays. Since we are in (or close to) the maximal stop-mixing regime, we expect from~\eq{eq:kappagstop} that the Higgs-boson production cross section should be suppressed with respect to the SM. In fact, we obtain $R_h \approx 0.95$ (and stop masses $m_{{\tilde t}_1} \approx 1.3 \, {\rm TeV}$ and $m_{{\tilde t}_2} \approx 1.6 \, {\rm TeV}$) throughout the entire parameter space depicted in the panels. The deviations in the relative signal strength $R_b$ of the Higgs-boson decay to bottom quarks are more pronounced than those in $R_h$ and correspond to enhancements of roughly~$20\%$. Notice that these shifts are due to the  terms $\epsilon_b$  and $( \Delta M_h^2 )_{\tilde t}$ in~(\ref{eq:kappab}), the latter of which introduces a positive correlation between $R_b$ and $M_h$. This feature clearly manifests itself in the two panels. As a consequence of the suppression of  $R_h$ and the enhancement of $R_b$, the Higgs-boson decays to $W$-, $Z$-boson, and tau pairs are all suppressed. For the considered MSSM parameters, we find $R_{W,Z} \approx 0.7$ and $R_\tau \approx [0.65,0.80]$. We add that $R_\tau$ shows a noticeable dependence on the higgsino mass parameter that stems from chargino and neutralino effects in the tau Yukawa coupling. While the structure of these corrections  is similar to those in~(\ref{eq:deltab}), these contributions have, for simplicity, not been included in the approximate formula~(\ref{eq:kappatau}). 

\begin{figure}[!t]
\begin{center}
\includegraphics[height=5cm]{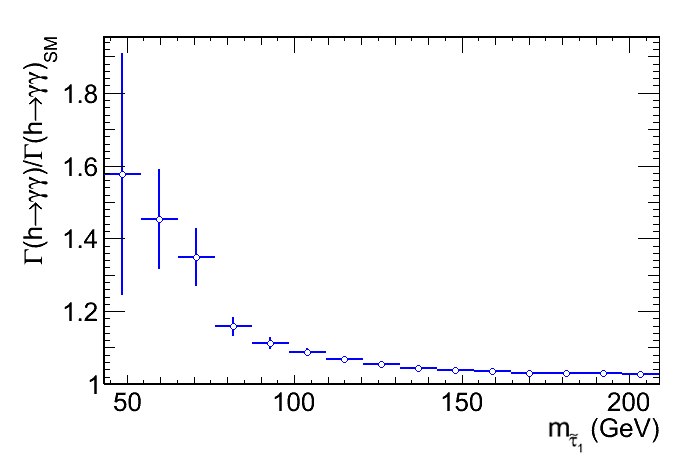} \qquad 
\raisebox{2mm}{\includegraphics[height=4.65cm]{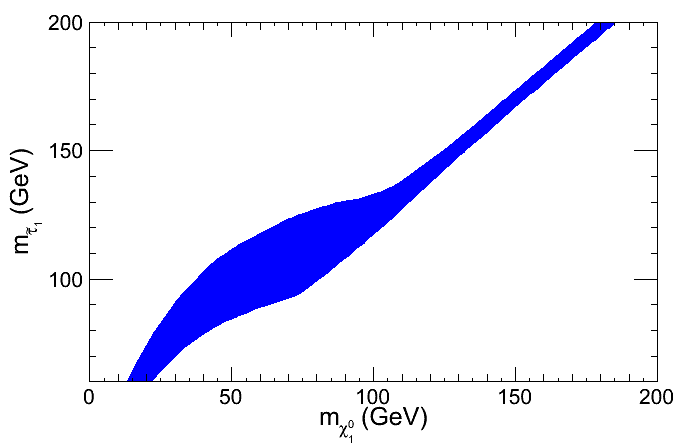}}
\end{center}
\vspace{-8mm}
\begin{center}
\parbox{15.5cm}{\caption{{Left: Average $h\to\gamma\gamma$ partial decay width normalised to its SM value as a function of the lighter stau mass. The shown vertical bars indicate the one standard deviations from the averaged values. Right: Values of the lighter stau mass necessary to obtain the correct DM abundance as a function of the lightest neutralino mass. In the parameter space below (above) the blue region  the predicted values of $\Omega_{\rm DM} h^2$ are below (above) the $3.5 \sigma$ WMAP bound.}\label{fig:2}}}
\end{center}
\end{figure}

As anticipated after~\eq{eq:kappagstau} and~\eq{eq:Rgamma}, the prediction for the  diphoton signal $R_\gamma$ depends very strongly on the amount of mixing $X_\tau \approx  \mu \, t_\beta$ in the stau sector. We observe that in the studied scenario, values of $R_\gamma > 1$ can only be obtained in a narrow mass window around $\mu \approx  900 \, {\rm GeV}$. For our choice of soft SUSY-breaking masses $\tilde m_{L_3} = \tilde m_{l_3} = 350 \, {\rm GeV}$, such large $\mu$  parameters lead to values of $m_{{\tilde \tau}_1}$ that are close to the LEP bound of $m_{{\tilde \tau}_1} > 81.9 \, {\rm GeV}$ \cite{Beringer:1900zz}. To further illustrate the latter feature we display in the left panel of Figure~\ref{fig:2} the maximal value for the $h \to \gamma\gamma$ decay width normalised to its SM value, that can be obtained for fixed $m_{{\tilde \tau}_1}$. In order to arrive at the plot we have used the results from a general 19 parameter scan in the phenomenological MSSM, as described in~\cite{Arbey:2011un,Arbey:2011aa}. The fast decoupling of the stau corrections in  $\Gamma( h \to \gamma\gamma)$ is evident from the figure.

We have seen that achieving a pronounced $h \to \gamma \gamma$ signal requires the presence of a large $\tilde \tau_1 \tilde \tau_1 h$ coupling. Such an interaction can be potentially problematic, since it may trigger additional minima in the scalar potential, and as a result the electroweak-breaking vacuum can become metastable~\cite{Rattazzi:1996fb,Hisano:2010re,Sato:2012bf,Kitahara:2012pb}. A tree-level analysis of the stability of the vacuum, taking into account large left-right mixing in the stau sector, leads to the following constraint~\cite{Hisano:2010re}
\beq \label{eq:vacuum}
 |\mu |  \hspace{0.25mm} t_\beta \lesssim 38.5 \left ( \sqrt{\tilde m_{L_3}} + \sqrt{\tilde m_{l_3}} \right )^2 - 10 \, {\rm TeV} \,.
\eeq 
Although this result cannot be fully trusted as loop effects are very important  in the specific MSSM scenarios considered in our work, it is clear from Figure~\ref{fig:Atmugrid} that in the slice of parameter space that features significant enhancements of  the diphoton signal, the bound (\ref{eq:vacuum})  is violated by around $30 \%$. In order to understand to which extend vacuum stability considerations exclude a light stau explanation of $R_\gamma \approx 1.7$,   a full one-loop analysis of  stau effects in the scalar potential would be required. Such a study is beyond the scope of this article.  

We start the discussion of our numerical results for the low-energy observables with $B \to X_s \gamma$. Adding to the  uncertainty of the SM prediction ${\Br} (B \to X_s \gamma)_{\rm SM} = (3.08 \pm 0.24) \cdot 10^{-4}$~\cite{Misiak:2006zs,Misiak:2006ab,Mahmoudi:2007vz} an intrinsic MSSM error of $0.10$ as well as the error of the experimental world average ${\Br} (B \to X_s \gamma)_{\rm exp} = (3.43 \pm 0.22) \cdot 10^{-4}$~\cite{Amhis:2012bh}, leads to the following $68\%$ confidence level~(CL) bound 
\beq \label{eq:RXsbound}
R_{X_s} = 1.11 \pm 0.11 \,,
\eeq
where the individual uncertainties have been added in quadrature. At $95\% \, {\rm CL}$ one obtains instead $R_{X_s} = [0.89, 1.33]$. The border of this 95\% probability region is indicated in the panels of Figure~\ref{fig:Atmugrid} by the dashed black curves and arrows.  From the panel showing $R_{X_s}$, one infers that for the depicted parameter choices, ${\Br} (B \to X_s \gamma)$ is always enhanced (by about 20\% to 60\%) with respect to the SM expectation. As expected from~(\ref{eq:C7chargino}), the  MSSM corrections grow with $A_t$ and become too large for $A_t\gtrsim4.5 \, {\rm TeV}$ to allow for an agreement with  ${\Br} (B \to X_s \gamma)$ at the $95\% \, {\rm CL}$.  This is an interesting and potentially important finding, since the parameter space disfavoured by $B \to X_s \gamma$ partially overlaps with that preferred by other observations/measurements. In fact, we see from the figure that the $B \to X_s \gamma$  constraint starts cutting into the already narrow regions in the $A_t$--$\mu$ plane with $M_h \approx 125 \,{\rm GeV}$ and $R_\gamma > 1$. 

A glimpse at the predictions for $R_{\mu^+ \mu^-}$ also shows that  there is an intriguing correlation between ${\Br} (B_s \to \mu^+ \mu^-)$ and ${\Br} (B \to X_s \gamma)$. Notice first that the obtained  branching fractions of $B_s \to \mu^+ \mu^-$ are all fully  compatible  with the  bound 
\beq \label{eq:Rmumubound}
R_{\mu^+ \mu^-} < 1.5 \,,
\eeq
which derives from the $95 \% \, {\rm CL}$ exclusion ${\Br} (B_s \to \mu^+ \mu^-)_{\rm exp} < 4.2 \cdot 10^{-9}$ \cite{Bsmumucombination} and the untagged SM branching fraction ${\Br} (B_s \to \mu^+ \mu^-)_{\rm SM} = 3.9 \cdot 10^{-9}$~\cite{Mahmoudi:2007vz,DeBruyn:2012wj,DeBruyn:2012wk} (corresponding to a CP-averaged branching ratio of $3.5  \cdot 10^{-9}$), after including a theoretical $68\% \, {\rm CL}$ error of~$20\%$ that is thought to cover both the SM and SUSY uncertainties. Second, we  observe that essentially all solutions that satisfy $B \to X_s \gamma$ feature a suppression of $B_s \to \mu^+ \mu^-$. In fact, asking for both an agreement with $R_{X_s}$ at the $95\% \, {\rm CL}$ as well as an enhanced diphoton rate, implies $R_{\mu^+ \mu^-} \approx [0.6, 1.0]$. Notice that given the $1/M_A^2$ dependence of  (\ref{eq:Rmumu}) and (\ref{eq:CP}) the ratio $R_{\mu^+ \mu^-}$ is a very sensitive measure of the masses of the heavy Higgses. It follows that the deviations quoted above can  be reduced by choosing $M_A \gg 1 \, {\rm TeV}$. An observation of the purely leptonic $B_s$ decay at the SM level (which seems possible with 2012 LHC data), may hence give important insights both on the nature of the $h \to \gamma \gamma$ excess in the context of the MSSM as well as the size of the decoupling scale $M_A$.

The last remaining low-energy observable in our study is the anomalous magnetic moment of the muon. Despite many changes and improvements in the recent history, the discrepancy seen in $a_\mu$ seems to persist. Combining the experimental result~\cite{Bennett:2006fi} with the SM calculation, based on an update of the hadronic vacuum polarisation contributions~\cite{Hagiwara:2011af} and the complete tenth-order QED corrections \cite{Aoyama:2012wk}, results in
\beq \label{eq:Deltaamu}
\Delta a_\mu = a_\mu^{\rm exp} - a_\mu^{\rm SM} = (2.5 \pm 0.9) \cdot 10^{-9} \,.
\eeq 
We see that for the  sneutrino masses $m_{\tilde \nu} \approx 1 \, {\rm TeV}$ present in our scenario, the predicted shifts in $a_\mu$ amount to about $[1.0, 1.6] \cdot 10^{-9}$, which leads to a significant reduction of the above tension.  The observed anti-correlation between $\Delta a_\mu$ and $\mu$ is readily understood from (\ref{eq:amuchargino}). For $\mu$, $m_{\tilde \nu} \gg M_2$,  the chargino-sneutrino corrections to the muon anomalous magnetic moment scale as $\Delta a_\mu^{\chi^\pm} \propto M_2/\mu \, M_W^2/m_{\tilde \nu}^2$. This relation also  implies  that a notable improvement in~(\ref{eq:Deltaamu})  requires  a relatively light slepton spectrum below a TeV, which in our scenario is present due to a suitable choice of parameters. The observed correlation between $R_\gamma > 1$ and  $\Delta a_\mu= {\cal O}(10^{-9})$ should therefore not be regarded as a solid prediction in the entire MSSM parameter space. We add that if slepton mass universality is assumed \cite{Giudice:2012pf}, the correlation between $R_\gamma$ and $\Delta a_\mu$  becomes however quite robust. 

\begin{figure}[!t]
\begin{center}
\includegraphics[width=0.32 \textwidth]{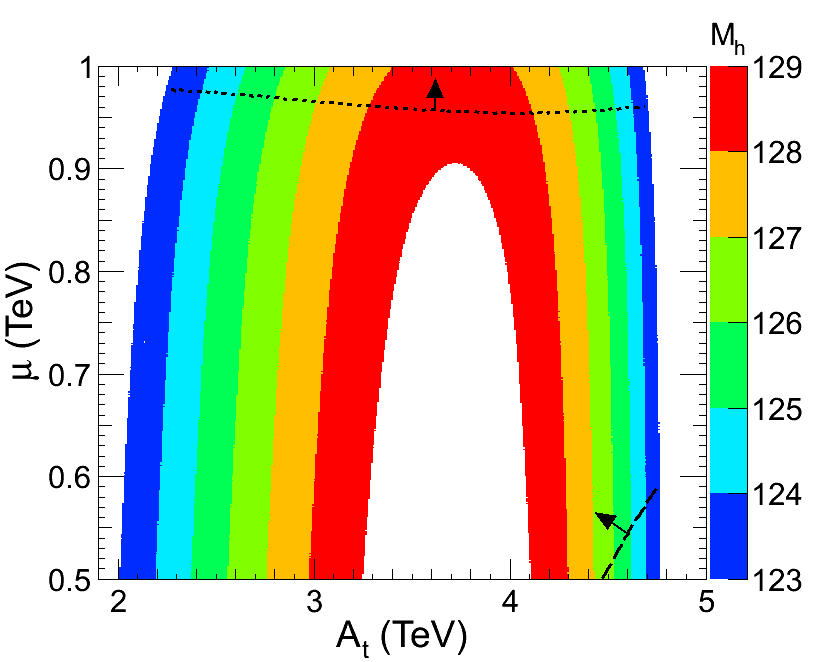}
\includegraphics[width=0.32 \textwidth]{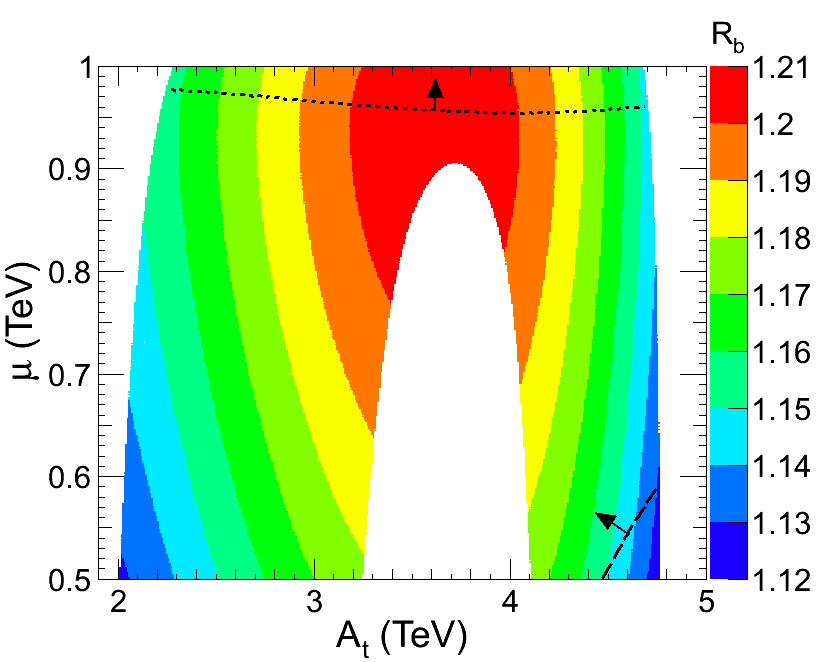}
\includegraphics[width=0.32 \textwidth]{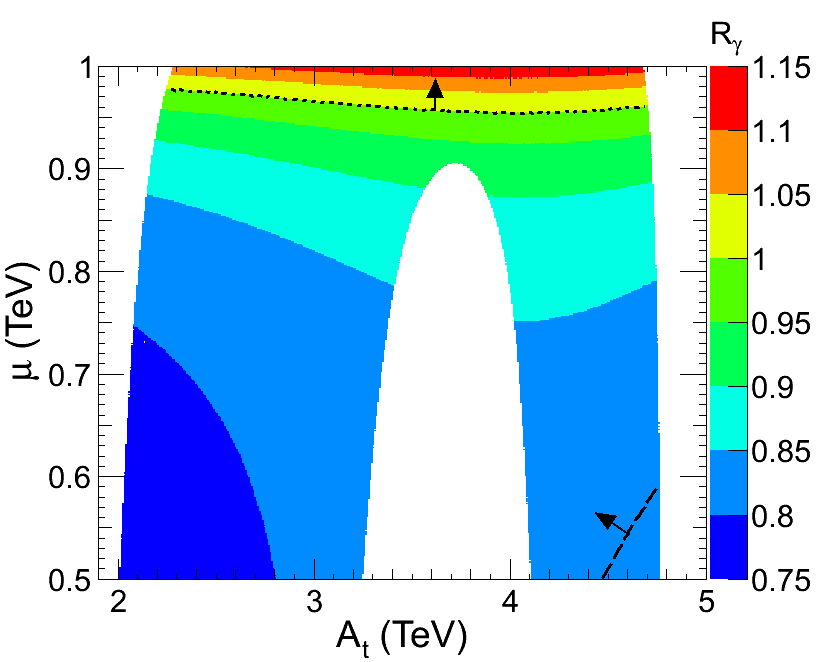}\\
\includegraphics[width=0.32 \textwidth]{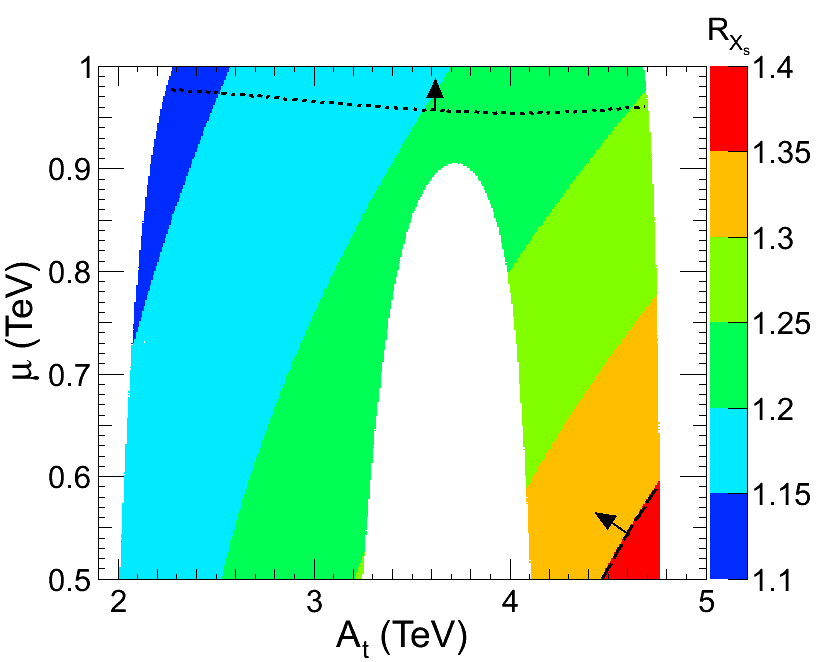}
\includegraphics[width=0.32 \textwidth]{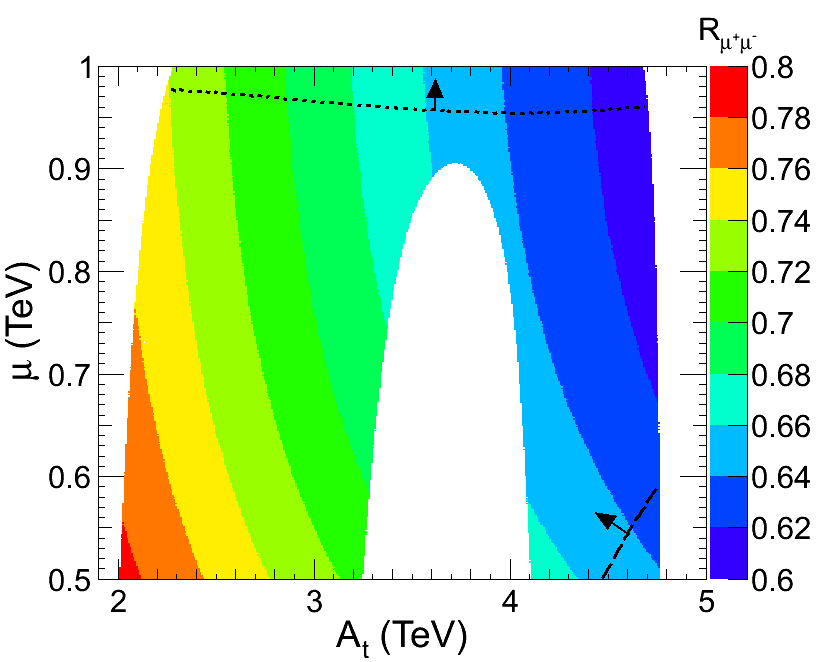}
\includegraphics[width=0.32 \textwidth]{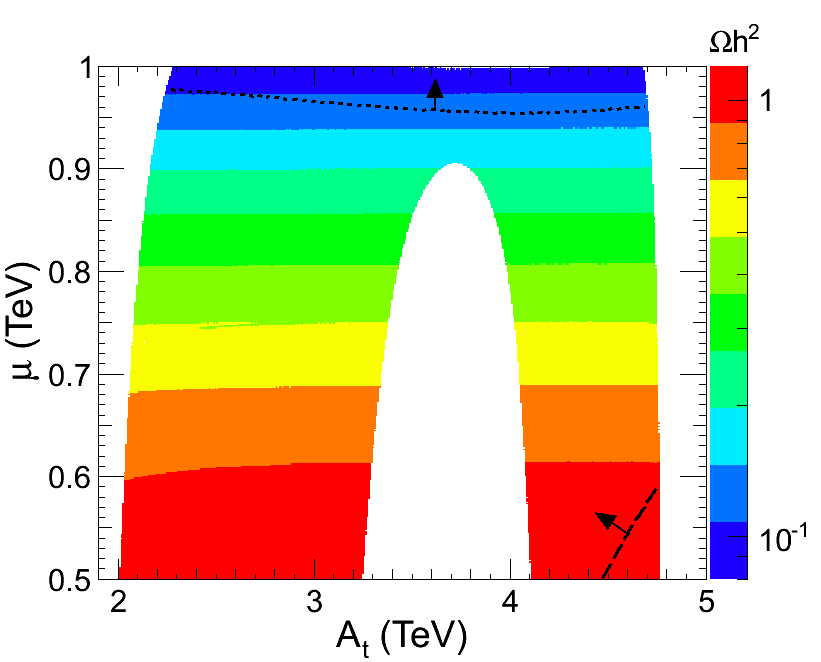}
\end{center}
\vspace{-6mm}
\begin{center}
\parbox{15.5cm}{
\caption{{Predictions for $M_h$ (upper left), $R_{b}$ (upper center), $R_{\gamma}$ (upper right), $R_{X_s}$ (lower left), $R_{\mu^+ \mu^-}$ (lower center), and $\Omega_{\rm DM} h^2$~(lower right) in the $t_\beta = 30$ scenario. The mass of the Higgs boson is given in units of $\rm GeV$. The dotted black lines indicate the parameter regions with $R_\gamma > 1$, while the dashed black lines correspond to the $95\% \,{\rm CL}$ regions favoured by $B \to X_s \gamma$. See text for further explanations.}\label{fig:Atmugridlowtb}}}
\end{center}
\end{figure}  

Let us now switch gear again and finally examine the predictions for the DM relic abundance. We see that the obtained values for $\Omega_{\rm DM} h^2$ range over three orders of magnitude, but that agreement with the tight WMAP $3.5 \sigma$  bound \cite{Komatsu:2010fb}
\beq \label{eq:WMAP}
\Omega_{\rm DM} h^2 = [0.068, 0.155] \,,
\eeq
that includes theoretical uncertainties~(see for example~\cite{Arbey:2012na} and references therein), can be achieved. In fact, requiring only that the LSP does not overpopulate the universe, \ie, $\Omega_{\rm DM} h^2 < 0.155$, singles out a parameter region in the $A_t$--$\mu$ plane that overlaps with that featuring $R_\gamma > 1$. The strong anti-correlation (positive correlation) between the $\Omega_{\rm DM} h^2$ and $\mu$ ($m_{{\tilde \tau}_1}$) is also clearly visible in the panels. It is easy to understand these two features by considering the pure annihilation  contribution~(\ref{eq:Omegah2bino}) to $\Omega_{\rm DM} h^2$ that effectively  limits the size of the mass splitting $m_{{\tilde \tau}_1} -m_{\chi_1^0}$. Numerically, we find that for our choice $M_1 = 50 \, {\rm GeV} \approx m_{\chi_1^0}$, the requirement of an electrically neutral LSP with  $\Omega_{\rm DM} h^2 < 0.155$ is only fulfilled if $m_{{\tilde \tau}_1} \approx [80, 120] \, {\rm GeV}$. Since $\tilde m_{L_3} = \tilde m_{l_3} = 350 \, {\rm GeV}$, such relatively light staus can however only be obtained for large $\mu$ parameters. The strong correlation between the mass of the LSP and the lighter stau is illustrated in the right panel of  Figure~\ref{fig:2}, which displays the parameter region in the $m_{\chi_1^0}$--$\hspace{0.5mm} m_{\tilde \tau_1}$ plane that is consistent with~(\ref{eq:WMAP}). The shown predictions correspond to the $t_\beta = 60$ scenario parameter choices with the value of $M_1$ varied. From the figure it is clear that for a fixed value of $m_{\tilde \tau_1}$ only a narrow range of $m_{\chi_1^0}$ values is consistent with the WMAP bound. This is turn implies that in the light stau scenario a confirmation of the excess in $pp \to h \to \gamma \gamma$ will have implication for direct and indirect DM searches, since the LSP mass is not a free parameter, but fixed to some degree.

The above discussion should have made clear that a confirmation of the results on the Higgs-boson couplings may point towards  rather peculiar (and technically ``unnatural'') MSSM parameters, namely large (and positive) values of $t_\beta$, $A_t$, and $\mu$.  These special parameter choices lead in turn to interesting and testable correlations between various observables. In the following, we would like to address the question of how robust these correlations are against 
the variations of  some of other MSSM parameters that have been  kept fixed so far.   

\begin{figure}[!t]
\begin{center}
\includegraphics[width=0.32 \textwidth]{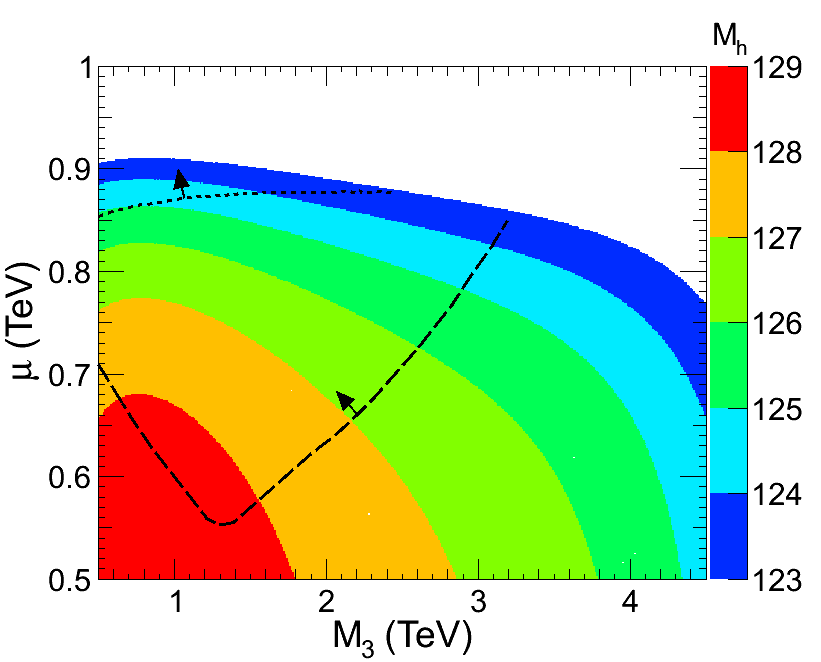}
\includegraphics[width=0.32 \textwidth]{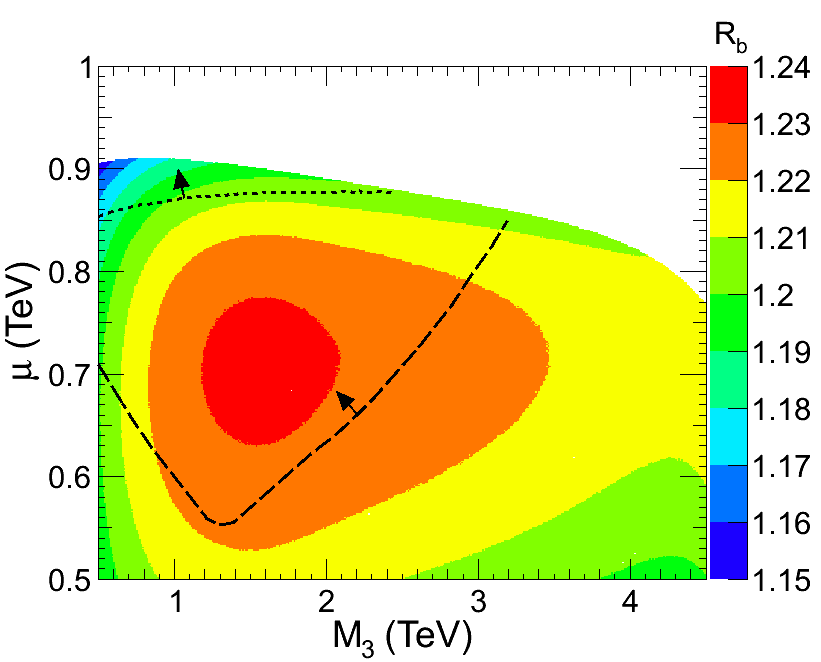}
\includegraphics[width=0.32 \textwidth]{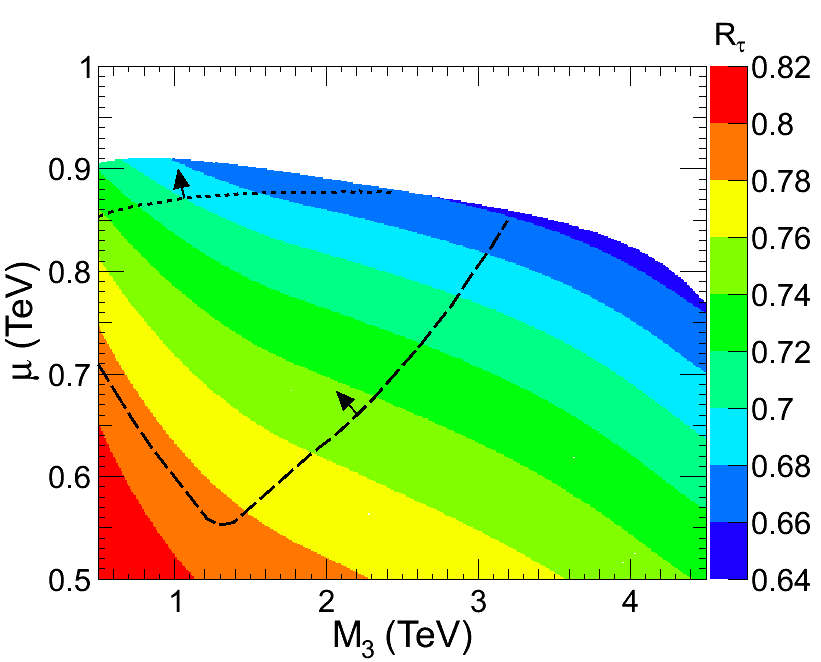}\\
\includegraphics[width=0.32 \textwidth]{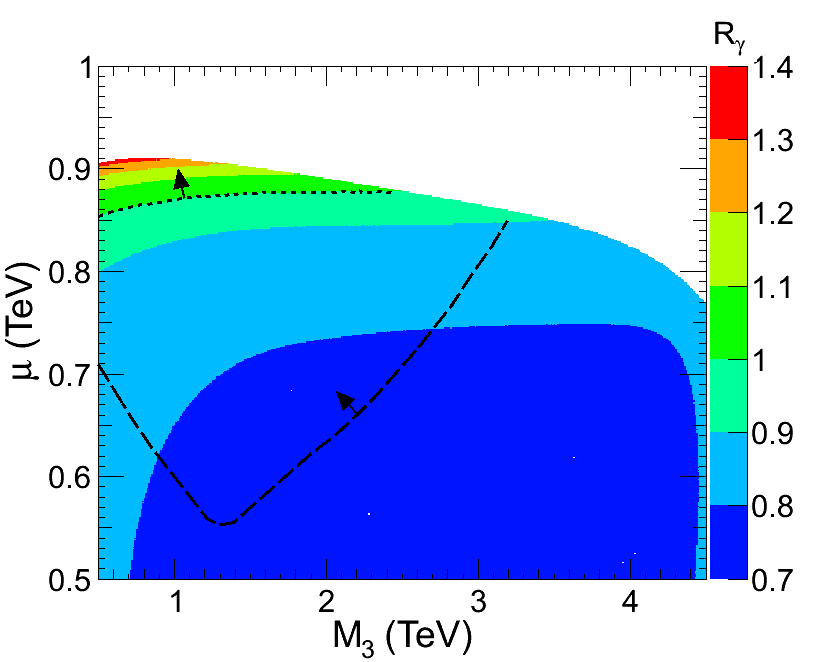}
\includegraphics[width=0.32 \textwidth]{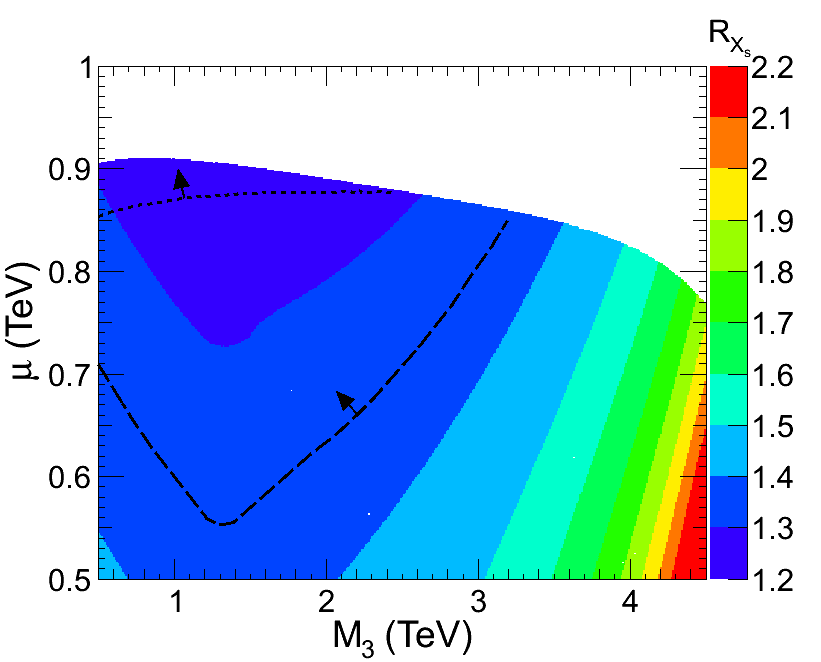}
\includegraphics[width=0.32 \textwidth]{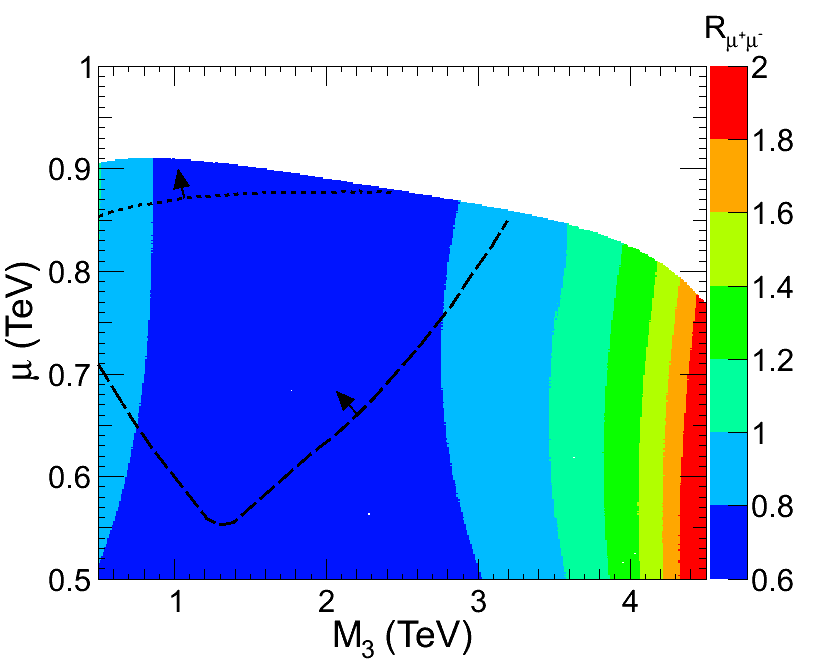}
\end{center}
\vspace{-6mm}
\begin{center}
\parbox{15.5cm}{
\caption{{Predictions for $M_h$ (upper left), $R_{b}$ (upper center), $R_{\tau}$ (upper right), $R_{\gamma}$ (lower left)), $R_{X_s}$ (lower center), and $R_{\mu^+ \mu^-}$ (lower right) in the $t_\beta = 60$ scenario for $A_t = 3$ TeV. The mass of the Higgs boson is given in units of $\rm GeV$. The dotted black lines indicate the parameter regions with $R_\gamma > 1$, while the dashed black lines correspond to the $95\% \,{\rm CL}$ regions favoured by $B \to X_s \gamma$.
See text for further explanations.}\label{fig:M3mugrid}}}
\end{center}
\end{figure}  

We start our discussion by studying the impact of $t_\beta$. In what  follows we employ $t_\beta = 30$, $\tilde m_{L_3} = 170 \, {\rm GeV}$, and $\tilde m_{l_3} = 350 \, {\rm GeV}$ (``$t_\beta = 30$ scenario''). Notice that the change of the soft SUSY-breaking masses is  required in order to obtain a very light $\tilde \tau_1$ eigenstate, which in turn results in a notable shift in $h \to \gamma \gamma$.  Furthermore, this choice of parameters satisfies the vacuum stability bound~(\ref{eq:vacuum}). The results of our numerical scans are collected in Figure~\ref{fig:Atmugridlowtb}. Let us first consider the Higgs-boson mass as well as the $h \to b \bar b$  and $h \to \gamma \gamma$ decay signals. We see that while the general pattern of the predictions resembles that obtained for $t_\beta = 60$, certain differences are clearly visible. First, now  even values of $\mu \gtrsim 1 \, {\rm TeV}$ lead to allowed Higgs-boson masses in the range $[123, 129] \, {\rm GeV}$. This feature is related to the negative corrections~(\ref{eq:Dmh2sb}) to the Higgs-boson mass that scale as $(\Delta M_h^2)_{\tilde b, \tilde \tau}  \propto t_\beta^4$. Second, the enhancements in $R_b$ are slightly smaller than those found for $t_\beta = 60$, but still amount to shifts in the ballpark of $20\%$. In turn, $R_{W,Z}$ and $R_\tau$ turn out to be somewhat larger than in the $t_\beta = 60$ scenario. Third, the enhancements in $R_\gamma$ are significantly smaller now and limited to~$15\%$.  This is expected since the stau corrections to the effective $h \to \gamma \gamma$ vertex~(\ref{eq:kappagstau})  scale  like $X_\tau^2 \propto t_\beta^2$. 

Turning to the predictions for $B \to X_s\gamma$, $B_s \to \mu^+ \mu^-$, and $\Omega_{\rm DM} h^2$, we first observe that in the $t_\beta = 30$ scenario, SUSY effects still tend to enhance the rate of the inclusive radiative $B$ decay, but that the corrections are about a factor of 2 smaller than before, \ie, $R_{X_s} \approx [1.1,1.4]$. Obviously, this is a result of the linear $t_\beta$ scaling of the chargino corrections~(\ref{eq:C7chargino}). From~(\ref{eq:amuchargino}) it follows that the same kind of depletion is also present  in $\Delta a_\mu$. In the case of $B_s \to \mu^+ \mu^-$, we find that in the parameter space favoured by $B \to X_s \gamma$, the branching ratio of the purely leptonic $B_s$ decay is always suppressed with respect to the SM expectation. One has $R_{\mu^+ \mu^-} \approx [0.6,0.8]$. This feature can again be understood from the interplay of $t_\beta$-enhanced terms in (\ref{eq:CP}). As in the case of $t_\beta = 60$, we finally see that acceptable values of the DM relic density can be achieved in the parts of the $A_t$--$\mu$ plane that also give $R_\gamma >1$.

\begin{figure}[!t]
\begin{center}
\includegraphics[width=0.32 \textwidth]{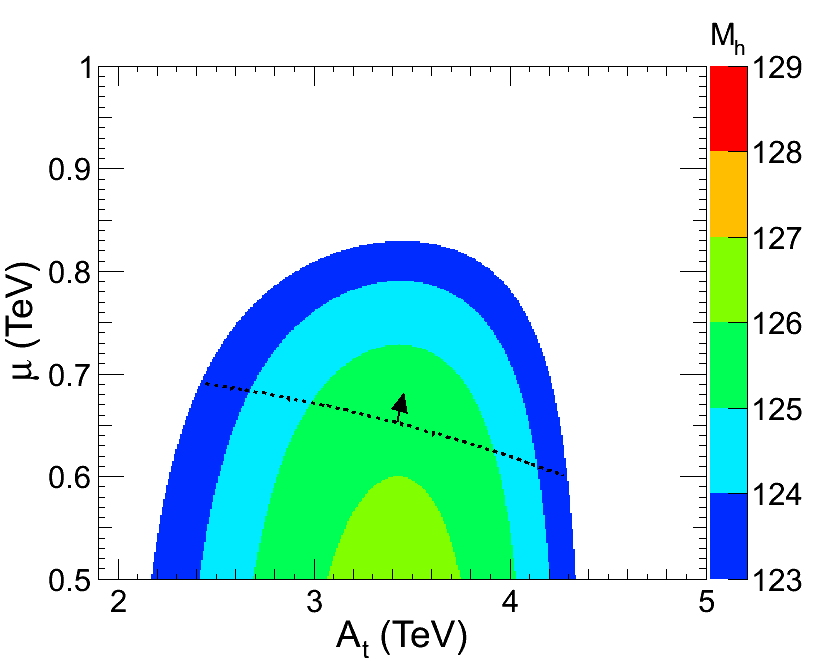}
\includegraphics[width=0.32 \textwidth]{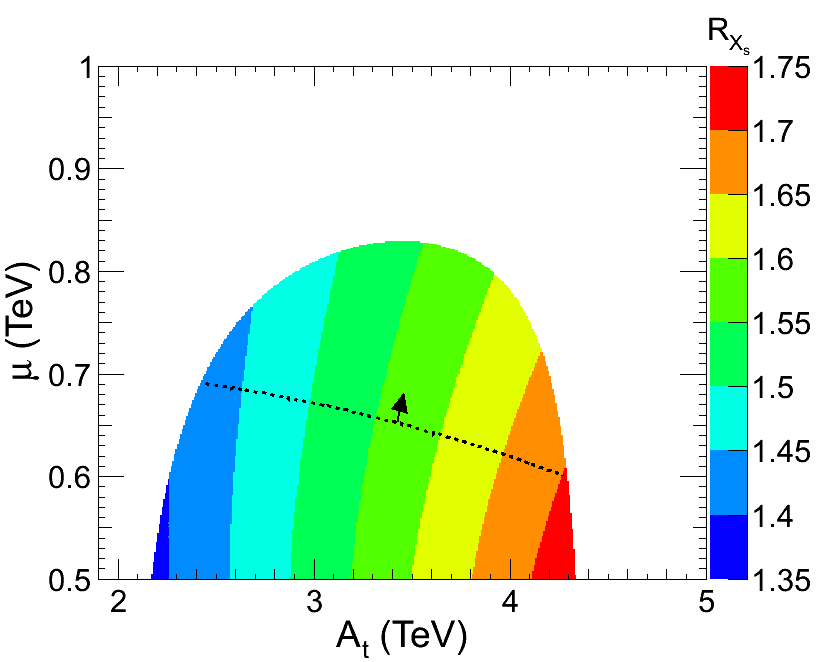}
\includegraphics[width=0.32 \textwidth]{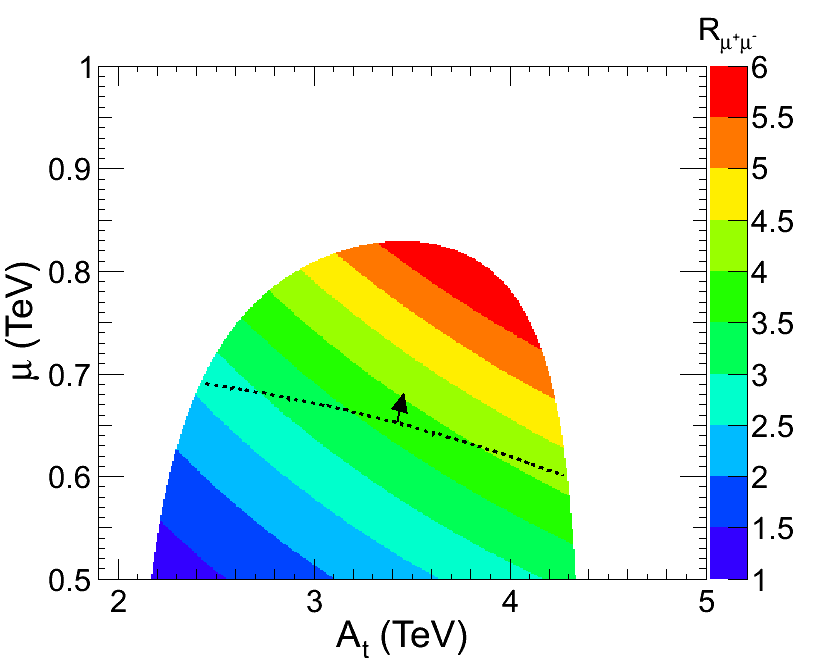}
\end{center}
\vspace{-6mm}
\begin{center}
\parbox{15.5cm}{
\caption{Predictions for $M_h$ (left), $R_{X_s}$ (center), and $R_{\mu^+ \mu^-}$ (right) in the $t_\beta = 60$ scenario, employing  $M_3 = -1.2 \, {\rm TeV}$. The mass of the Higgs boson is given in units of $\rm GeV$.  The dotted black lines indicate the parameter regions with $R_\gamma > 1$. See text for additional details.\label{fig:AtmugridM3negative}}}
\end{center}
\end{figure}  

Another important SUSY parameter is the gluino mass $M_3$. In order to study its impact, we perform scans in the $\mu$--$M_3$ plane, fixing the value of the trilinear stop-Higgs boson coupling to $A_t = 3 \, {\rm TeV}$ and employing the parameters of the $t_\beta = 60$ scenario discussed before. Our most important findings are illustrated in the panels of  Figure~\ref{fig:M3mugrid}. In the case of the Higgs-boson mass we observe that for fixed $\mu$ parameter the predictions for $M_h$ reach a maximum for $M_3 \approx 1 \, {\rm TeV}$ and then start decreasing for increasing gluino masses. This effect is associated to the one-loop gluino corrections  (\ref{eq:deltab}) to the bottom  Yukawa coupling that lead to a negative shift in $M_h$ via~(\ref{eq:Dmh2sb}). In fact, we infer from the  predictions for $R_\gamma$ that the most pronounced enhancement in this observable  occur in a thin stripe with $\mu \approx 900 \, {\rm GeV}$ and $M_3 \approx [0.5, 1.5] \, {\rm TeV}$. Requiring $M_h > 123 \, {\rm GeV}$  and $R_\gamma > 1$ hence effectively sets an upper limit on the gluino mass.  In the slice of the $\mu$--$M_3$ plane that leads to an enhanced $h \to \gamma \gamma$ signal, we also see that the predictions for $h \to b\bar b$, $h \to \tau^+ \tau^-$, $R_{X_s}$, and $R_{\mu^+ \mu^-}$ vary only moderately with $M_3$. Numerically, we find that $R_b \approx [1.15,1.21]$, $R_\tau \approx [0.66,0.76]$, $R_{X_s} \approx [1.2,1.3]$, and $R_{\mu^+ \mu^-} \approx [0.6,1.0]$.  Notice that $R_b$ increases with $M_3$, while $R_\tau$ and $R_{\mu^+ \mu^-}$ both decrease with increasing gluino mass.  These behaviours can be traced back to the $t_\beta$-enhanced gluino corrections entering (\ref{eq:deltab}). The observables not explicitly  displayed in the figure show either essentially no ($R_h$,  $\Delta a_\mu$, and $\Omega_{\rm DM} h^2$) or only a minor  dependence ($R_{W,Z}$) on the gluino mass. 

So far we have restricted ourselves to the case of a positive gluino mass term $M_3$. In order to see whether this sign choice has some impact on the obtained results, we repeat our numerical scans in the $t_\beta = 60$ scenario, using $M_3 = -1.2\, {\rm TeV}$ instead of $M_3 = 1.2 \, {\rm TeV}$. The corresponding plots can be found in Figure~\ref{fig:AtmugridM3negative}. From the predictions for $M_h$, we see that for $M_3 < 0$, Higgs-boson masses above $127 \, {\rm GeV}$ cannot be achieved. This feature is related to the fact that for $\mu M_3 < 0$ the tree-level bottom and tau Yukawa couplings are enhanced with respect to the case of $\mu M_3 > 0$, which leads to larger negative corrections (\ref{eq:Dmh2sb}) to $M_h$.  We add that the Higgs-boson decays to massive gauge bosons and fermions all remain essentially SM-like in the studied scenario, while $R_\gamma$ can be enhanced by up to 35\%. For what concerns the $B$-physics observables, we observe that flipping the sign of $M_3$ while leaving the remaining MSSM parameters untouched, enhances both $\Br (B \to X_s \gamma)$ and $\Br (B_s \to \mu^+ \mu^-)$. Notice that these enhancements originate from the $t_\beta$-enhanced gluino corrections appearing in (\ref{eq:deltab}).  While in the case of $B \to X_s \gamma$ the predicted values of the branching ratio are larger by around a factor of $1.1$,  we find that the branching fraction of $B_s \to \mu^+ \mu^-$ is increased by factors $2.5$ to $5$. Given the stringent bounds (\ref{eq:RXsbound}) and (\ref{eq:Rmumubound}), scenarios with $M_3 < 0$ and large $t_\beta$, $A_t >0$, and $\mu >0$ are hence disfavoured.  The remaining observables do not significantly depend on the sign chosen for $M_3$, and we hence do not show the corresponding predictions in the figure.

From the above explorations in the MSSM parameter space, we conclude that many of the found correlations are  robust, as long as one restricts oneself to the region with $t_\beta \gtrsim 50$, large and positive $A_t$ and $\mu$, $M_h \approx 125 \, {\rm GeV}$, and an enhanced $h \to \gamma \gamma$ rate. Positive values of $M_3$ are also clearly favoured over negative gluino mass parameters. 

Let us finally add that a strong enhancement of $R_\gamma$ can also be achieved by suppressing the partial decay rate of the Higgs boson to bottom pairs. While this can be easily achieved in the MSSM, a suppression of $R_b$ leads typically to enhanced $h \to WW, ZZ$ rates. Given that the ratios $R_{W,Z}$ appear to be SM-like~\cite{ATLAS:2012gk, CMS:2012gu} an explanation of $R_\gamma \approx 1.7$ via a suppressed $h \to b\bar b$ width is (at present) not favoured by experiment.

\section{Conclusions}
\label{sec:conclusions}

The announcements of the discovery of a  bosonic state by the LHC high-$p_T$ experiments mark the beginning of a new chapter in particle physics. While the significance of the various measurements is not yet sufficient to tell if  the properties of the observed particle agree with that of the SM Higgs scalar, the preliminary findings of an enhanced $h \to \gamma \gamma$ rate have triggered a lot of excitement, in particular, in the theoretical community.   With ATLAS and CMS accumulating more data, the question of whether new physics or just a statistical fluctuation is responsible for the observed deviation, may be answered by the end of this year.

In  this article we have studied in detail under which circumstances MSSM scenarios with a light stau can give rise to a significant enhancement of the diphoton signal without violating other existing constraints  from  $B$ physics, $(g-2)_\mu$, and dark matter. We found that the observation of a Higgs-like state with a mass of around  $125 \, \GeV$  combined with the preliminary measurements of the Higgs-boson couplings points towards a distinct (but unnatural) choice of parameters, namely large values of $\tan \beta$,  $M_A$, $A_t$, and $\mu$ with ${\rm sgn} \left (A_t \right ) = {\rm sgn}\left  (\mu \right ) = +1$. In this region of parameter space the correct  thermal relic density can be achieved, but only if one assumes the hierarchy $|M_1| \ll |M_2| \ll |\mu|$.  A typical MSSM spectrum leading to a significantly enhanced $h \to \gamma \gamma$ rate as well as the correct value of $\Omega_{\rm DM} h^2$, therefore contains a light bino as the dark matter candidate, a light and maximally mixed stau (often causing  the vacuum to become metastable),  and a heavy higgsino. We also showed that spectra where the gluino is  much heavier than the squarks can be problematic, since in such a case $|\mu|$ and in turn the enhancements in $h \to \gamma \gamma$ are bounded from above.  

In the corner of phase space singled out by the Higgs-boson mass, the diphoton rate, and the relic density, we found that the predictions for the remaining Higgs-boson observables are relatively robust against variations of the other MSSM parameters.  In fact, intriguing patterns of deviations surface. While Higgs-boson  production is typically slightly suppressed with respect to the SM,  the $h \to b\bar b$ rate is generically enhanced by around~20\%, which in turn results in  suppressions of  the decays to $W$-boson, $Z$-boson, and tau pairs  by a comparable amount. Such shifts are in good agreement with the tentative findings  by the LHC and Tevatron collaborations on the search for the SM Higgs scalar. 

A further consequence of the large and positive values of $\tan \beta$, $M_A$, $A_t$, and $\mu$ are $B \to X_s \gamma$ branching fractions that are above the SM expectation by about~$30\%$. In view of the ongoing effort to improve the theoretical understanding of the inclusive radiative $B$ decay, deviations of such an amount may provide a smoking gun signal of the light stau scenario in the future. Let us add that in contrast to $B \to X_s \gamma$, the predictions for $B \to X_s l^+ l^-$, $B \to K^\ast l^+ l^-$, and $B \to \tau \nu$ all turn out to be SM-like and well in agreement with experimental results. Similarly, the values of the $B \to D^{(\ast)} \tau \nu$ branching ratios are very close to the SM expectations. The anomalies  seen in the latter channels \cite{Lees:2012xj} can hence not be accommodated in the light stau scenarios (nor in the full MSSM). 
We also observe that in the region of parameter space favoured by $B \to X_s \gamma$ the rate of $B_s \to \mu^+ \mu^-$ tends to be smaller than the SM prediction. 
 The sign of the gluino mass parameter plays a crucial role in obtaining viable predictions for the observables in the $B$-meson sector and we found that positive values of $M_3$ are clearly favoured over parameter choices with $M_3 <0$. Since the corrections in  $B_s \to \mu^+ \mu^-$  can reach up to~40\%  (for not too large values of $M_A$), precision measurements of the purely leptonic $B_s$ decay, now under way at ATLAS, CMS, and LHCb, might shed further light on whether the observed enhancement in $h \to \gamma \gamma$ is due to a light stau.  We finally showed that under the assumption of light  soft-SUSY breaking slepton masses, the long-standing discrepancy in $(g-2)_\mu$ is significantly reduced in  the parameter region selected by the Higgs-boson mass and the enhanced diphoton signal. 

The choice of the scenarios studied in this paper allowed us to identify the correlations which would have been otherwise difficult to highlight in a general MSSM scan. These correlations  might become very useful in case the ATLAS and CMS experiments continue to measure a notable enhancement of $h \to \gamma \gamma$ and/or start to see the first supersymmetric partners. 

\subsubsection*{Acknowledgements}
\label{subsec:acknowledgments}

We are grateful to Ben~Allanach, Thomas~Hahn, Sven~Heinemeyer, Sabine~Kraml, and Michael Spira for useful discussions and correspondence. FM would like to thank Alexandre~Arbey and Marco~Battaglia for useful discussions and comments on the manuscript. UH would like to thank Babis~Anastasiou for an invitation to  ``The Zurich Phenomenology Workshop, Higgs Search Confronts Theory'', which triggered the present work. He also acknowledges travel support from the UNILHC network (PITN-GA-2009-237920). FM acknowledges partial support from the European Union FP7 ITN INVISIBLES (Marie Curie Actions, PITN-GA-2011-289442).

\end{document}